\documentclass[12pt]{article}
\pdfoutput=1
\usepackage{myjheppub}
\usepackage{graphicx,booktabs,latexsym,amssymb,amsmath,bm,epsfig,psfrag,float,placeins}
\hypersetup{pdfnewwindow=true}

\preprint{
\begin{flushright}	

	TUM-HEP-1148/18,\\
\end{flushright}}

\title{\boldmath Master Integrals for double real radiation emission in heavy-to-light quark decay \unboldmath}

\author[a,b]{Roberto Bonciani,}
\author[c]{Alessandro Broggio,}
\author[d]{Leandro Cieri,}
\author[e,f]{and Andrea Ferroglia,}

\affiliation[a]{Sapienza - Universit\`a  di Roma, Dipartimento di Fisica, Piazzale Aldo Moro 5, 00185, Rome, Italy}
\affiliation[b]{INFN Sezione di Roma, Piazzale Aldo Moro 2, 00185, Rome, Italy}
\emailAdd{roberto.bonciani@roma1.infn.it}

\affiliation[c]{Physik Department T31, Technische Universit\"at M\"unchen,
	James Franck-Stra{\ss}e 1, D-85748 Garching, Germany}
\emailAdd{alessandro.broggio@tum.de}

\affiliation[d]{INFN, Sezione di Milano-Bicocca, Piazza della Scienza 3, 20126 Milan, Italy}
\emailAdd{cieri@mib.infn.it}

\affiliation[e]{Physics Department, New York City College of Technology, The City
	University of New York, 
	300 Jay Street, Brooklyn, NY 11201 USA}

\affiliation[f]{The Graduate School and University Center,
	The City University of New York, 365 Fifth Avenue,
	New York, NY 10016  USA}
\emailAdd{aferroglia@citytech.cuny.edu}

\abstract{We evaluate analytically the master integrals for double real radiation emission 
in the $b \to u W^*$ decay, where $b$ and $u$ are a massive and massless quark, respectively, 
while $W^{*}$ is an off-shell charged weak boson. Since the $W$ boson can subsequently decay 
in a lepton  anti-neutrino pair, the results of the present paper constitute a further step 
toward a fully analytic computation of differential distributions for the semileptonic decay 
of a $b$ quark at NNLO in QCD. The latter partonic process plays a crucial role in the study 
of inclusive semileptonic charmless decays of $B$ mesons. Our results are expressed in terms 
of multiple polylogarithms of maximum weight four.}

\begin{document}
\newcommand{\Red}[1]{\textcolor{red}{#1}}
\newcommand{\Green}[1]{\textcolor{green}{#1}}
\newcommand{\Blue}[1]{\textcolor{blue}{#1}}
\newcommand{\Cyan}[1]{\textcolor{cyan}{#1}}
\newcommand{\Magenta}[1]{\textcolor{magenta}{#1}}
\newcommand{\alert}[1]{{\bf \color{red}[{#1}]}}

\newcommand{\be}{\begin{equation}}
\newcommand{\ee}{\end{equation}}

\newcommand{\nn}{\nonumber}
\def\ff{f\hspace{-0.3cm}f}
\def\mgamc{{{\tt \small MG5\_aMC}}}

\maketitle

\section{Introduction}
\label{sec:intro}

The study of inclusive semileptonic $B$ meson decays is important for the determination
of the Cabibbo-Kobayashi-Maskawa (CKM) matrix elements $|V_{ub}|$ and $|V_{cb}|$ and,
therefore, it constitutes a stringent test on the unitarity of the CKM matrix.

Inclusive determinations of $|V_{ub}|$ rely on an Operator Product Expansion (OPE) \cite{Bigi:1992su,Blok:1993va,Manohar:1993qn,Gremm:1996df}, according to which 
the total $B$ meson semileptonic decay rate and various kinematic distributions can be described, at leading order in
a power expansion with respect to the inverse $b$-quark mass, in terms of the partonic decay rate of an on-shell $b$ quark
into a lepton-neutrino pair and a $u$ quark. Within this framework, theoretical predictions for  the partonic decay rates are obtained by means of perturbation theory. Phenomenological predictions for observables related to the semileptonic $B$-meson decays are then obtained by combining perturbative calculations for the semileptonic  $b$-quark decays with a finite number of non-perturbative parameters.

However, the measurements of the $B \to X_u e \bar{\nu}$ decay are affected by large backgrounds due to the $B \to X_c e \bar{\nu}$ decay. In order to suppress this background, experiments impose sharp cuts (for example, cuts on the final state hadronic invariant mass). This in turn leads to problems with the convergence of the OPE in theoretical predictions. These issues can be addressed by parameterizing the residual motion of a $b$-quark in the $B$ meson by means of the shape function \cite{Neubert:1993um,Neubert:1993ch,Bigi:1993ex}. Further studies \cite{Bauer:2001rc,Bauer:2000xf} showed that with a combination of cuts on the hadronic and leptonic invariant masses the impact of the shape function can be suppressed and the OPE can be used to describe the $B \to X_u e \bar{\nu}$. In both approaches the QCD corrections to the partonic process $b \to u W^*$ (where $W^*$ indicates an off shell $W$ boson and the mass of the up quark is set to zero) play a crucial role. 

An analytic result for the next-to-leading order (NLO) QCD corrections to the $b \to l \bar{\nu}_{l} u$ differential decay rate was obtained in \cite{DeFazio:1999ptt}. 
 A fully differential calculation of this  decay rate at next-to-next-to-leading order (NNLO) 
was carried out in  \cite{Brucherseifer:2013cu,Brucherseifer:2013iv}. A related  result for the differential top-quark semileptonic decay at NNLO was presented in \cite{Gao:2012ja}. These NNLO calculations are
based on numerical techniques. Analytic results at NNLO are also known; however so far these studies were carried out 
in the shape-function region by using Soft-Collinear Effective Theory  
\cite{Bell:2008ws,Asatrian:2008uk,Beneke:2008ei}. 
The contribution of these corrections to $|V_{ub}|$ were then considered in \cite{Greub:2009sv}.

In this paper, we focus on the analytic calculation of the Master Integrals (MIs) necessary for the determination
of the contribution of the QCD double-real radiation to the $b \to u W^*$ decay. This process is one of the three elements that are necessary 
for the evaluation of the triple-differential  distribution in the charged lepton energy, leptonic invariant mass and final-state hadronic invariant mass at NNLO in QCD. For what concerns the other two elements, the two-loop corrections to the $b \to u W^*$ decay were evaluated analytically in 
\cite{Bonciani:2008wf,Asatrian:2008uk,Bell:2008ws,Beneke:2008ei}, while the one-loop real-virtual contribution will be the subject of future work.

In order to carry out the calculation, we employed the method of  reverse unitarity. This approach was introduced in the context of the evaluation
of the NNLO QCD corrections to the production of a Higgs boson in gluon fusion \cite{Anastasiou:2002yz}
and then applied to several other  processes (see for example 
\cite{Melnikov:2004bm,Melnikov:2005bx,Asatrian:2006sm,Asatrian:2010rq,Anastasiou:2015ema,Bonciani:2016wya}). The method consists in applying Cutkosky rules \cite{Cutkosky:1960sp} in order to map the calculation of the interference between two leading order (LO) $2\to 3$ diagrams integrated over the final state phase-space into the evaluation of ``cuts'' of two-loop $2 \to 2$ diagrams. In this way, the on-shell condition for the real particles in the final state of the $2 \to 3$ process is converted
into the difference of two propagators with opposite $i0^+$ prescription. 
Subsequently, one can calculate the cut diagrams by means of techniques for the analytic evaluation of multi-loop diagrams which were developed starting from the late '90s. 
In particular, dimensionally regularized scalar integrals are reduced to  MIs by using Integration by-Parts
Identities (IBPs)
\cite{Tkachov:1981wb,Chetyrkin:1981qh,Laporta:2001dd}. Reduction algorithms based on IBPs are implemented in publicly available computer programs \cite{Anastasiou:2004vj,Lee:2012cn,Lee:2013mka,Maierhoefer:2017hyi,Smirnov:2008iw,Smirnov:2013dia,Smirnov:2014hma,Studerus:2009ye,vonManteuffel:2012np}. The MIs are analytically evaluated 
by means of the Differential Equations (DE) method \cite{Kotikov:1990kg,Remiddi:1997ny,Gehrmann:1999as,Argeri:2007up,Henn:2014qga} and expressed in terms of generalized polylogarithms
(GPLs) \cite{Goncharov:polylog,Goncharov2007,Remiddi:1999ew} of two variables: 
$t$, which is connected to the invariant mass of the hadronic final state, and $z$, which is 
related to the leptonic invariant mass. The evaluation of the $\epsilon$ expansion of the MIs was carried out up to terms that include GPLs of maximum weight four.

Our results are also relevant for the determination of the total width of a top quark that decays
into a massless bottom quark and a lepton-neutrino pair \cite{Gao:2012ja}.

The paper is structured as follows. In Section \ref{sec:calculation}, we discuss the calculation, by introducing the notation and the kinematics of the process, and by identifying  the MIs and the range of validity 
of their analytic expressions. In Section \ref{sec:results}, we present the results for the MIs.
Section \ref{sec:conclusions} contains our conclusions. Appendix \ref{sec:poles} collects the explicit 
expressions of the $\epsilon$-poles of the MIs.

The analytic expressions of the twelve MIs evaluated in this work are collected in an ancillary file 
included in the {\tt arXiv} submission.

\section{Calculation}
\label{sec:calculation}

The calculation of double emission corrections to the $b \to u W^*$ process is first mapped into the 
problem of calculating three-particle cuts in two-loop $bW^* \to bW^*$ forward box-diagrams, using the method proposed in \cite{Anastasiou:2002yz}.  
Three auxiliary topologies which encompass all of the combinations of denominators which can appear 
in the cut diagrams were subsequently identified. The MIs belonging to each topology were identified 
by using IBPs as implemented in {\tt LiteRed} \cite{Lee:2012cn,Lee:2013mka}. 
The MIs, which depend on two dimensionless parameters (defined below), are then calculated by employing 
the DE  method. The technique employed in this work is by now a standard method in the analytic calculation of Feynman diagrams. In this section we describe the way in which we parameterized the kinematics of the process, we define the MIs which were identified and we discuss the way in which the integration constants arising in the DE method were fixed.   
 
\subsection{Kinematics}

\begin{figure}[t]
	\begin{center}
		\begin{tabular}{cc}
			\includegraphics[width=7.2cm]{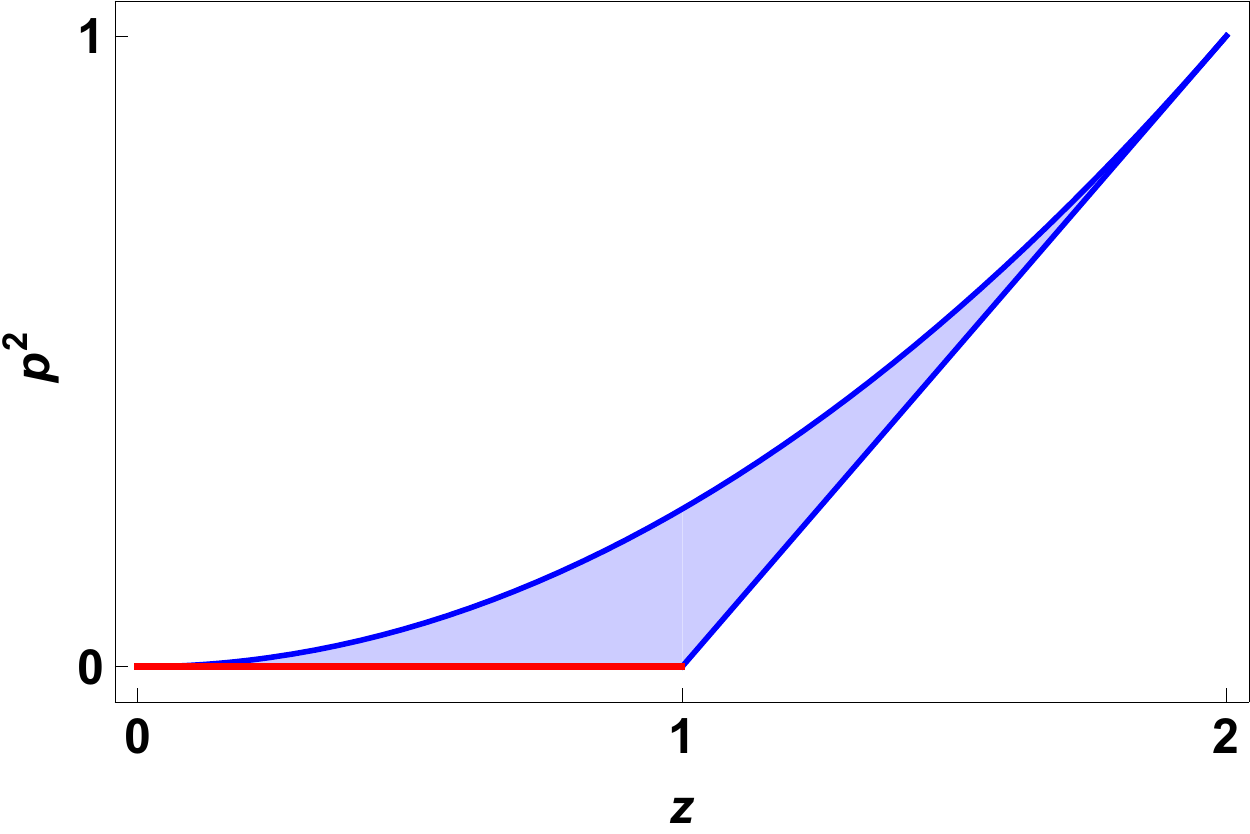} &  \includegraphics[width=7.2cm]{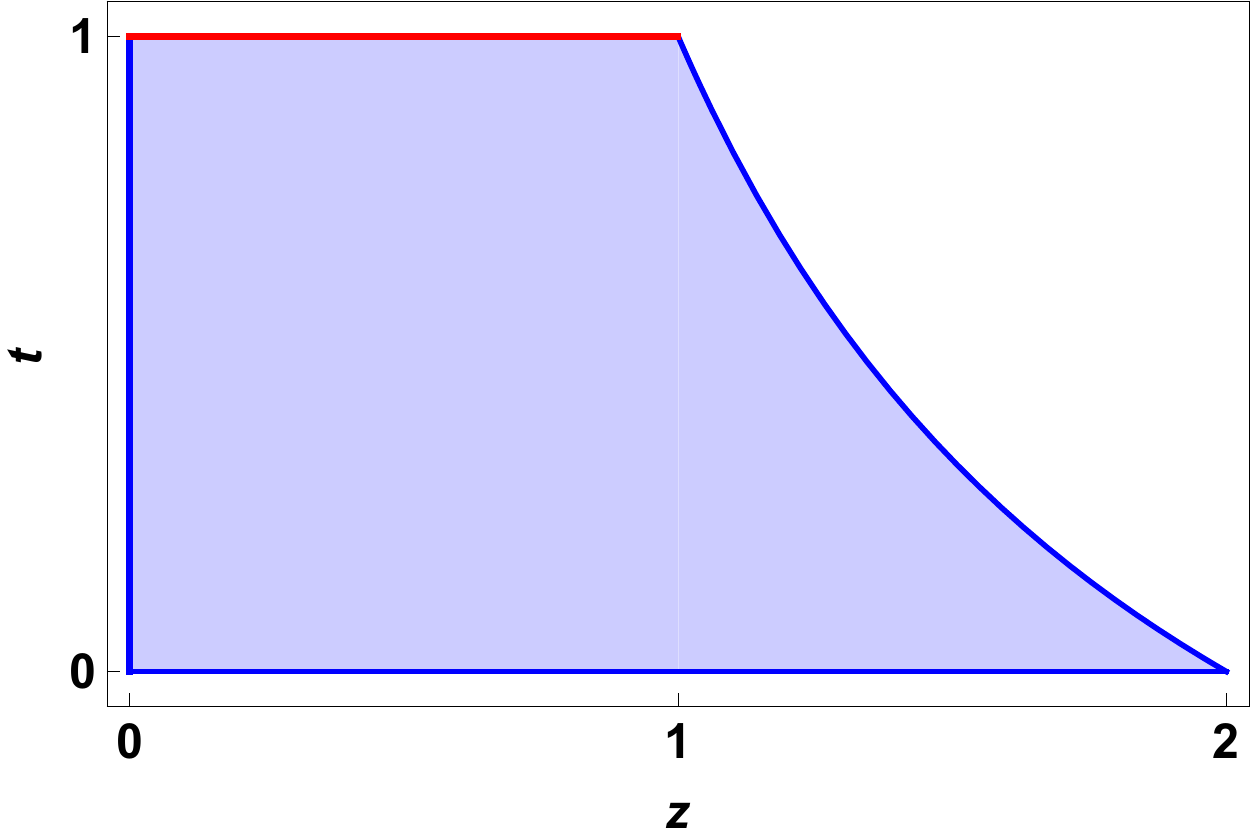}
		\end{tabular}
	\end{center}
	\caption{Physical phase space regions in the ${z,\hat{p}^2}$ plane (left panel) and ${z,t}$ plane (right panel). In both cases, the line corresponding to tree level kinematics is indicated in red. \label{fig:phasespace}}
\end{figure}

At tree-level, the kinematics of the decay we are interested in is 
\begin{equation}
b (p_1) \to W^-(q) + u (p_2)  \, , \qquad p_1 = p_2 +q \,,
\end{equation} 
where $p_1$, $p_2$ and $q$ are the four-momenta of the bottom quark, up quark and W boson, respectively.
Consequently, 
\begin{equation}
p_1^2 = m_b^2 \, , \qquad p_2^2 = 0 \, ,
\end{equation}
where $m_b$ indicates the mass of the bottom quark and the mass of the up quark is neglected. 
The $W$ boson is off-shell.
At tree level the energy of the up quark in the bottom-quark rest frame is smaller that $m_b/2$. Therefore, we introduce the dimensionless parameter $z$ defined as
\begin{equation}
z \equiv \frac{2 p_1 \cdot p_2}{m_b^2} \, , \qquad 0\le z \le 1 \, .
\label{eq:z}
\end{equation}

Beyond leading order, it is necessary to consider the process 
\begin{equation}
b (p_1) \to  W^-(q) + X(p_X) \, , \qquad p_1 = p_X +q \,,
\end{equation} 
where $X$ indicates an inclusive state involving light quarks and gluons, so that $p_X^2 \neq 0$ in general. As it was done in \cite{DeFazio:1999ptt}, the invariant mass of the state $X$ is parameterized  by introducing  the variable $t$ defined through the relation
\begin{equation}
p_X^2 \equiv m_b^2 \hat{p}^2 \equiv \frac{m_b^2}{4} z^2 \left(1 -t^2\right) \, ,
\end{equation}
where $z$ remains defined as in Eq.~(\ref{eq:z}), provided that $p_2$ is replaced by $p_X$.
Tree-level kinematics, i.e. $p_X^2 = 0$ with $0 \le z \le 1$, is recovered in the $t \to 1$ limit. Beyond tree level, the available phase space in the $\{\hat{p}^2,z\}$ and $\{t,z\}$ plane is shown in Figure~\ref{fig:phasespace}.
The physical region in the  $\{\hat{p}^2,z\}$ plane is shown in the left panel of Figure~\ref{fig:phasespace} and it is delimited by the conditions
\begin{equation}
0 \le z \le 2 \, , \qquad \max \{0,z-1\} \le \hat{p}^2 \le \frac{z^2}{4} \, . 
\end{equation}
In the $\{t,z\}$ plane the physical region (shown in the right panel of Figure~\ref{fig:phasespace})  is given by 
\begin{equation}
0 \le z \le 2 \, , \qquad   0 \le t \le \min \left\{1,\frac{2}{z}-1 \right\} \, . 
\end{equation}

In the calculation of the MIs which we carry out in this paper, we keep $t \neq 1$. 
In fact, the differential distribution which we ultimately want to obtain by employing 
the integrals which we evaluate here is divergent for $t \to 1$ and  includes ``star'' 
distributions of the form\footnote{Star distributions can be defined through the relation
\begin{displaymath}
\int_0^{\hat{m}^2} d \hat{p}^2 f\left(\hat{p}^2\right) \left(\frac{\ln^n \hat{p}^2}{\hat{p}^2}\right)_* = f(0) \frac{\ln^{n+1}{\hat{m}^2}}{n+1} + \int_0^{\hat{m}^2} d \hat{p}^2 \frac{\ln^n \hat{p}^2}{\hat{p}^2} \left[f\left(\hat{p}^2\right) -f(0)\right]\, ,
\end{displaymath}	
where $f$ is a smooth test function.
	} \cite{DeFazio:1999ptt}
\begin{displaymath}
\left(\frac{\ln^n \hat{p}^2}{\hat{p}^2}\right)_* \, , \qquad n = 0,1,2,3 \,.
\label{eq:star}
\end{displaymath}
However, it is sufficient to calculate the MIs for real radiation corrections by keeping 
$t \neq 1$ since the expression of the partonic double differential distribution in the 
$t \to 1$ limit was evaluated by using Soft Collinear Effective Theory (SCET) in 
\cite{Greub:2009sv}. The partonic real radiation corrections in the $t \to 1 \,\, (\hat{p}^2 \to 0)$ 
limit were recalculated also by us. These corrections, which receive contributions from double real and real virtual diagrams,   involve poles in $\epsilon$ and finite terms 
proportional to $\delta(\hat{p}^2)$ as well as the terms proportional to the star distributions of
Eq.~(\ref{eq:star}). If one neglects the star distribution bracket, the star distributions arising 
in the SCET calculation must match exactly the singular terms arising from the calculation carried 
out at $t \neq 1$. The latter depends on the MIs evaluated here as well as on the two cut MIs which 
will be the subject of a future work. Therefore, at the stage in which the differential distribution 
will be assembled, the SCET calculation of the real radiation will serve as a further cross check 
of the calculation carried out at $t \neq 1$. At the same time, it will provide the complete 
contribution of the real corrections proportional to $\delta(\hat{p}^2)$. We already tested with success 
this procedure by recalculating at NLO the differential distribution originally derived in \cite{DeFazio:1999ptt}.

\subsection{Auxiliary topologies and Master Integrals}

\begin{figure}[p!]
	\begin{center}
		\begin{tabular}{cc}
			\vspace*{-2.5cm} & \\
			\hspace*{-1cm}	\includegraphics[width=11.2cm]{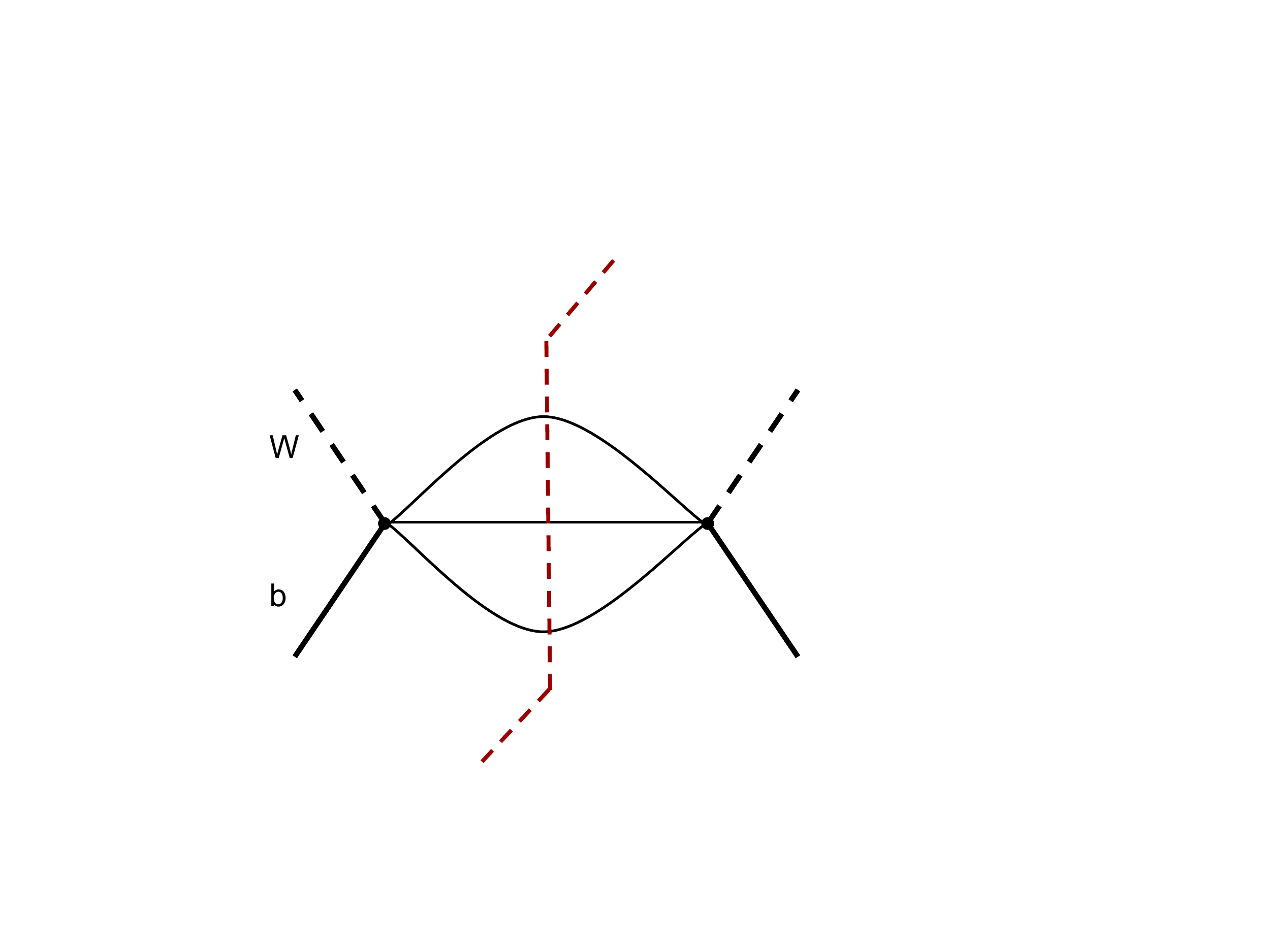} & \hspace*{-3.8cm} \includegraphics[width=11.2cm]{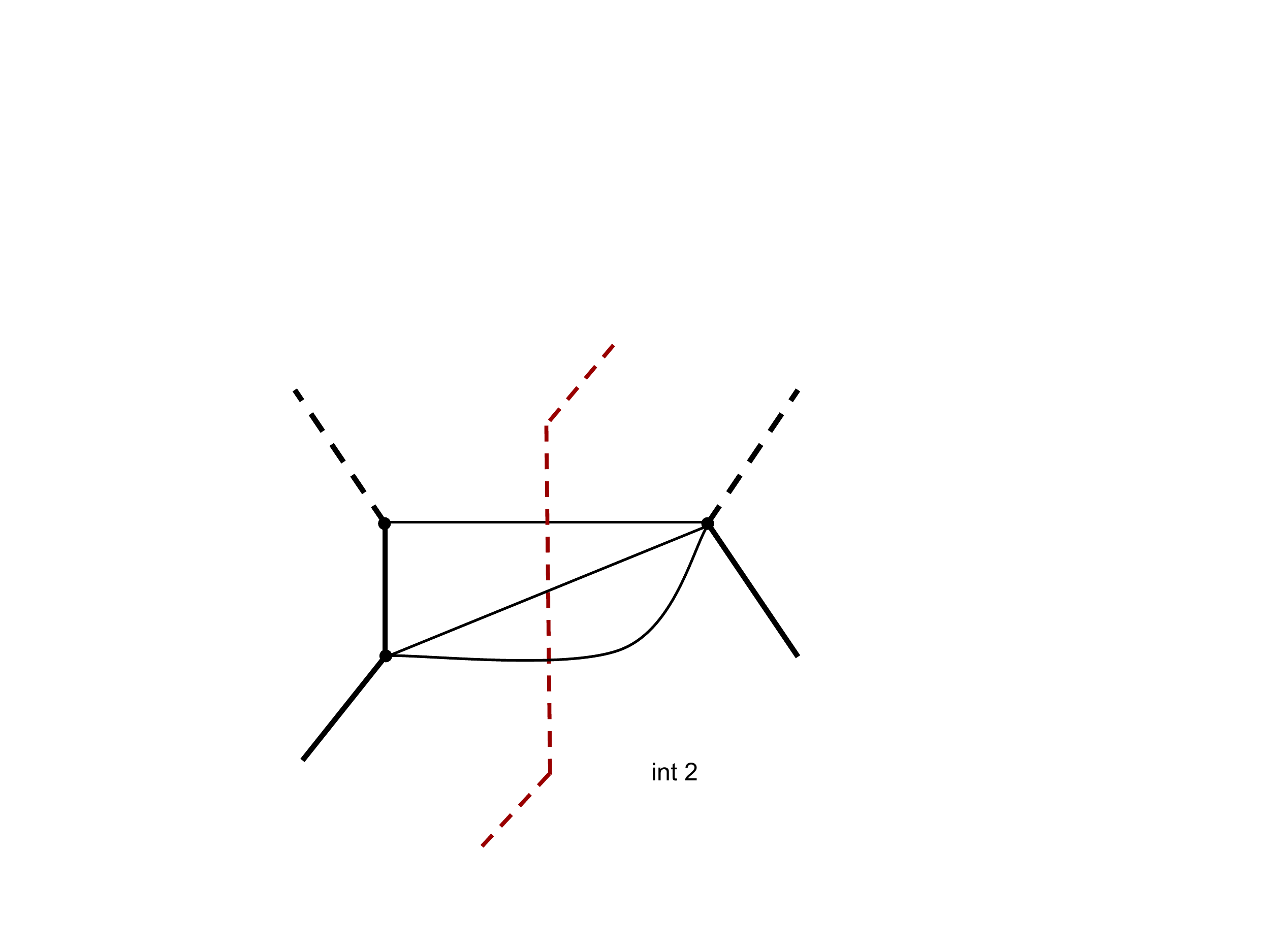} \\
			\vspace*{-4.5cm} & \\
			\hspace*{-1cm}	\includegraphics[width=11.2cm]{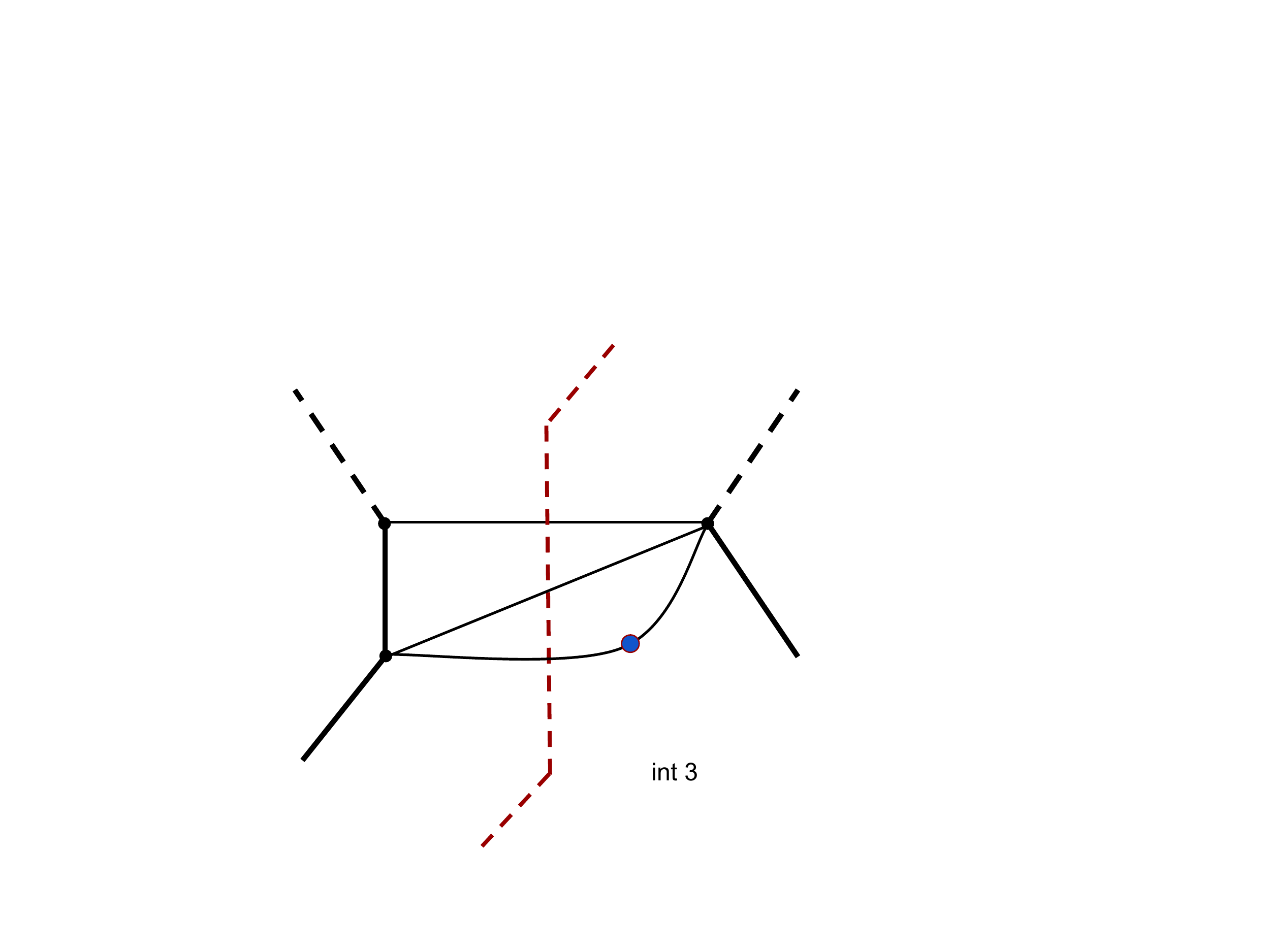} & \hspace*{-3.8cm} \includegraphics[width=11.2cm]{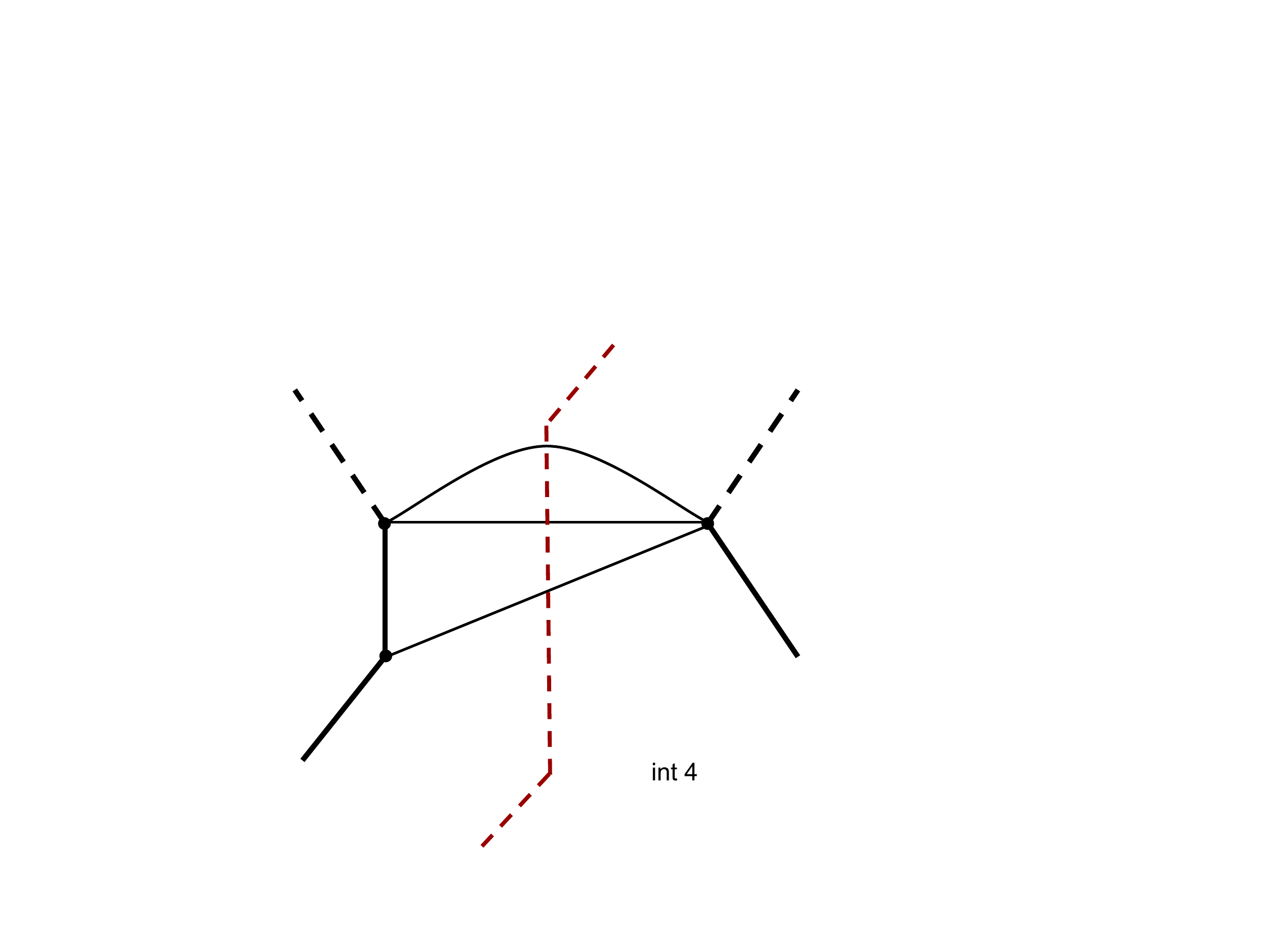} \\
			\vspace*{-4.5cm} & \\
			\hspace*{-1cm}	\includegraphics[width=11.2cm]{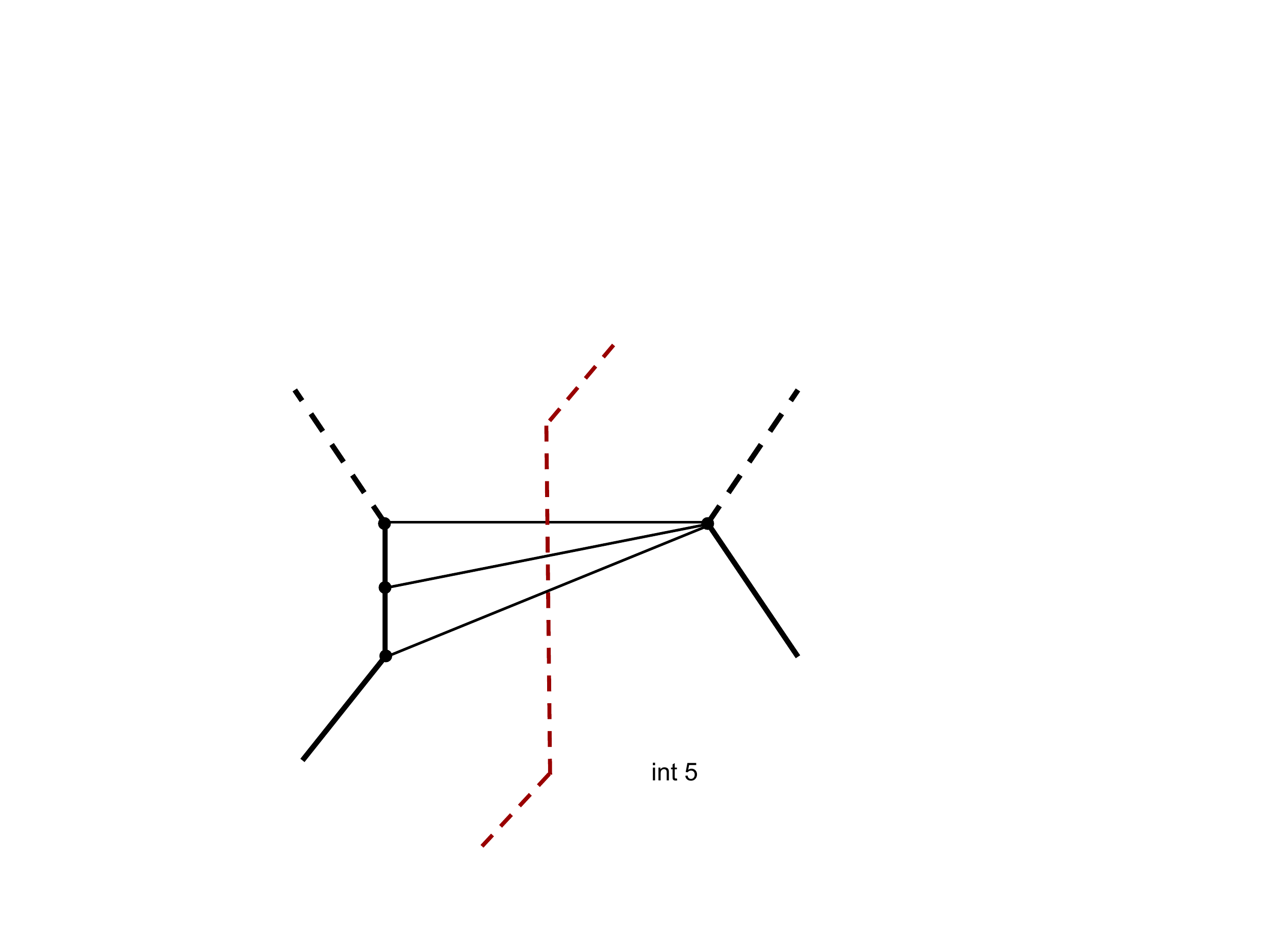} & \hspace*{-3.8cm} \includegraphics[width=11.2cm]{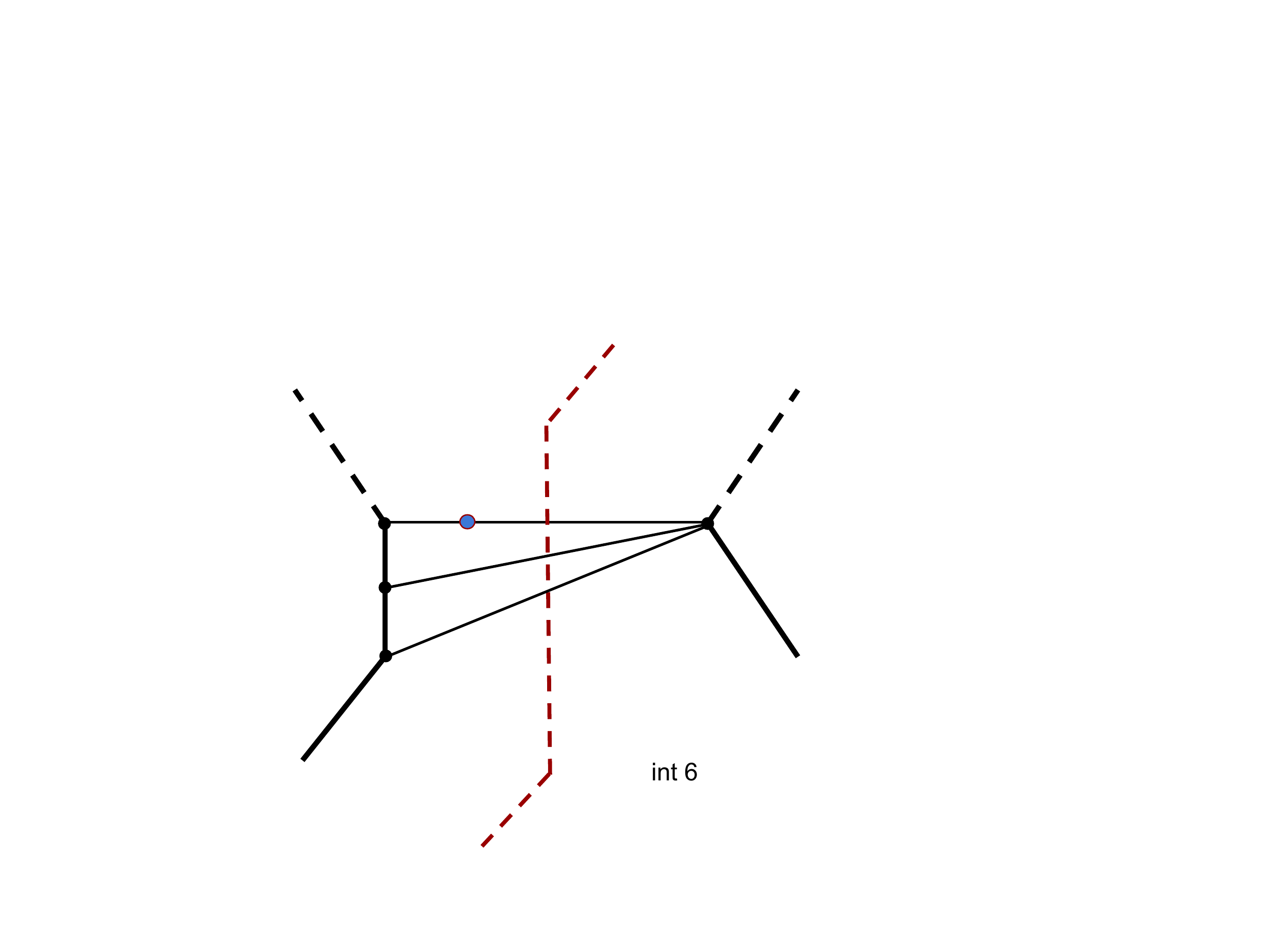} \\
			\vspace*{-4.5cm} & \\
			\hspace*{-1cm}	\includegraphics[width=11.2cm]{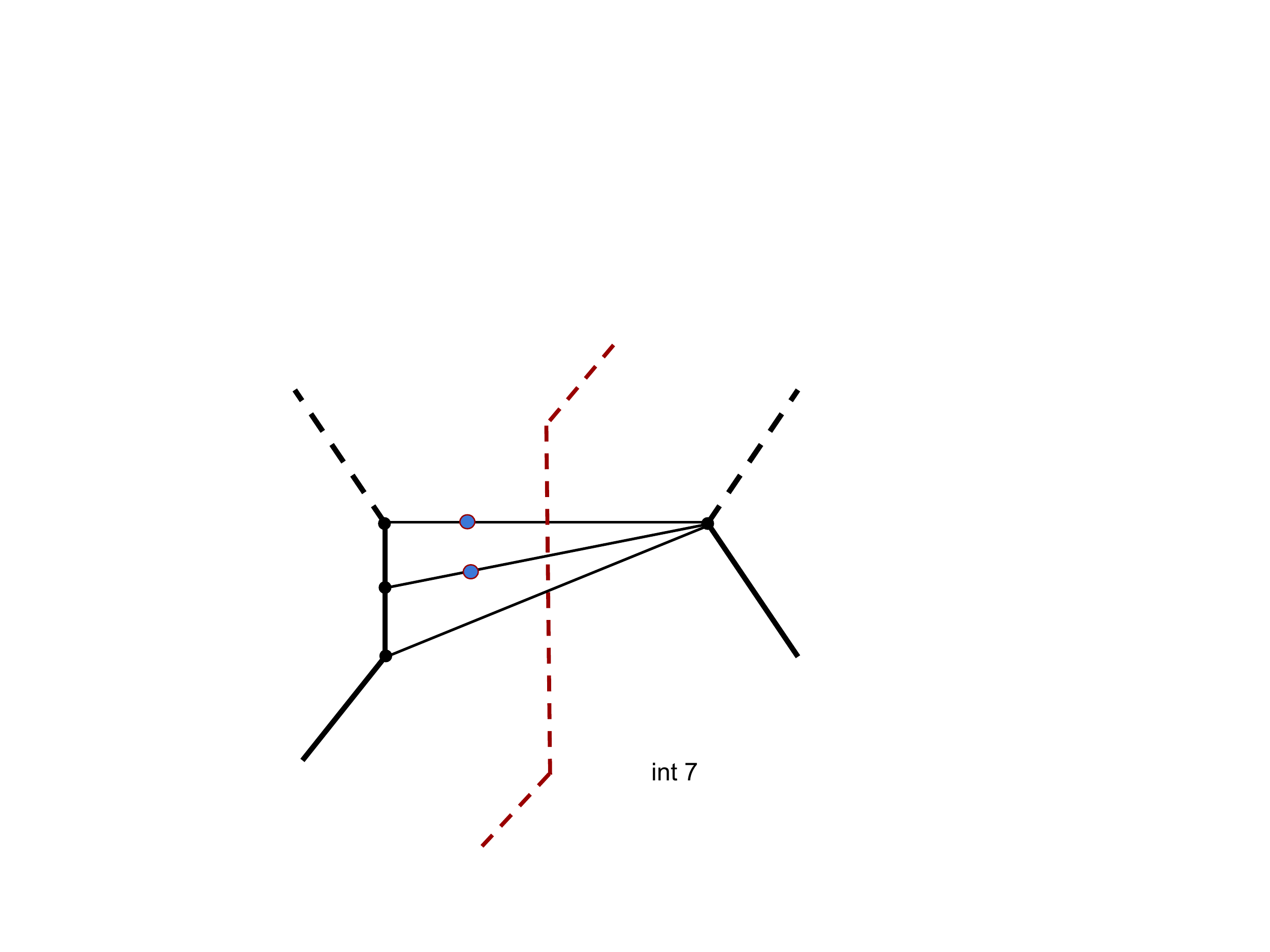} & \hspace*{-3.8cm} \includegraphics[width=11.2cm]{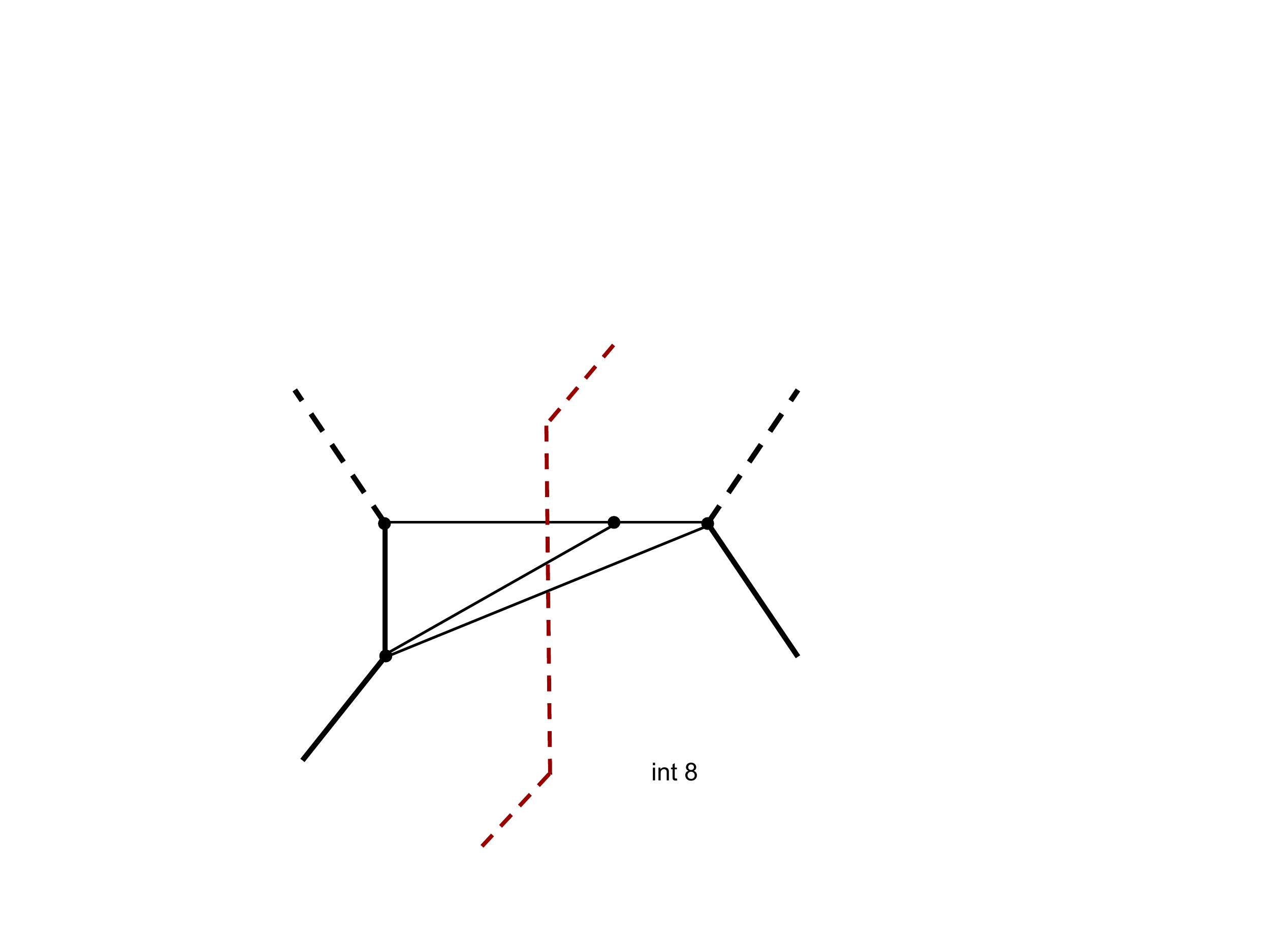} \\
		\end{tabular}
		\vspace*{-1cm}
	\end{center}
	\caption{Master Integrals for Topology A: Integrals $I_1, \cdots, I_8$. \label{fig:TopAMIS}}
	
\end{figure}

All of the Feynman diagrams contributing to the double emission corrections to the $b \to u W^*$ 
process can be calculated once the scalar integrals belonging to the (three cuts of the) three auxiliary topologies are known.

The integrals belonging to the first auxiliary topology considered, which is referred to as topology A, are defined as follows
\begin{equation}
I_A \left(\alpha_1, \alpha_2, \alpha_3, \alpha_4,\alpha_5, \alpha_6, \alpha_7 \right) = \int \frac{d^d k_1}{(2 \pi)^d} \frac{d^d k_2}{(2 \pi)^d} \prod_{i=1}^7  \frac{1}{P_i^{\alpha_i}} \, .
\label{eq:topA}
\end{equation}
The seven propagators $P_i$ in  Eq.~(\ref{eq:topA}) are 
\begin{align}
P_1 &=  k_1^2 \, , \quad
P_2 = (p_1 - k_1)^2 -m_b^2  \, , \quad
P_3 =(p_1 - k_1 - k_2)^2 -m_b^2 \, ,\quad
P_4 = (p_2 - k_1 - k_2)^2 \, , \nonumber \\
P_5 &= k_2^2\, ,\qquad
P_6 = (k_1+k_2)^2 \, \qquad
P_7 = (p_2-k_1)^2 \, ,
\label{eq:propA}
\end{align}
where the last three propagators in the list are cut propagators. For example
\begin{equation}
\frac{1}{P_5} \to \delta (k_2^2) = \frac{1}{2 \pi i} \left[ \frac{1}{k_2^2 +i 0^+} - \frac{1}{k_2^2 -i 0^+}  \right] \, .
\end{equation}
Equivalent relations hold for $P_6$ and $P_7$.
As a consequence of the presence of cut propagators, the integrals $I_A$ in Eq.~(\ref{eq:topA}) are zero when at least one among the powers $\alpha_5, \alpha_6, \alpha_7$ is zero or negative. The eight MIs belonging to topology A are shown in Figure~\ref{fig:TopAMIS}. We introduce the following notation in order to label the MIs of topology A:
\begin{align}
I_1 \equiv I_A \left(0,0,0,0,1,1,1 \right) \, , &\qquad   &I_2 \equiv I_A \left(0,1,0,0,1,1,1 \right)\, ,  \nonumber \\
I_3 \equiv I_A \left(0,1,0,0,1,2,1 \right) \, , &\qquad   &I_4 \equiv I_A \left(0,0,1,0,1,1,1 \right)\, ,  \nonumber \\
I_5 \equiv I_A \left(0,1,1,0,1,1,1 \right)\, ,  &\qquad   &I_6 \equiv I_A \left(0,1,1,0,1,1,2 \right) \, , \nonumber \\
I_7 \equiv I_A \left(0,1,1,0,2,1,2 \right)\, ,  &\qquad   &I_8 \equiv I_A \left(0,1,0,1,1,1,1 \right)\, .
\end{align}

\begin{figure}[ht!]
	\begin{center}
		\begin{tabular}{cc}
			\vspace*{-2.5cm} & \\
			\hspace*{-1.cm}	\includegraphics[width=11.2cm]{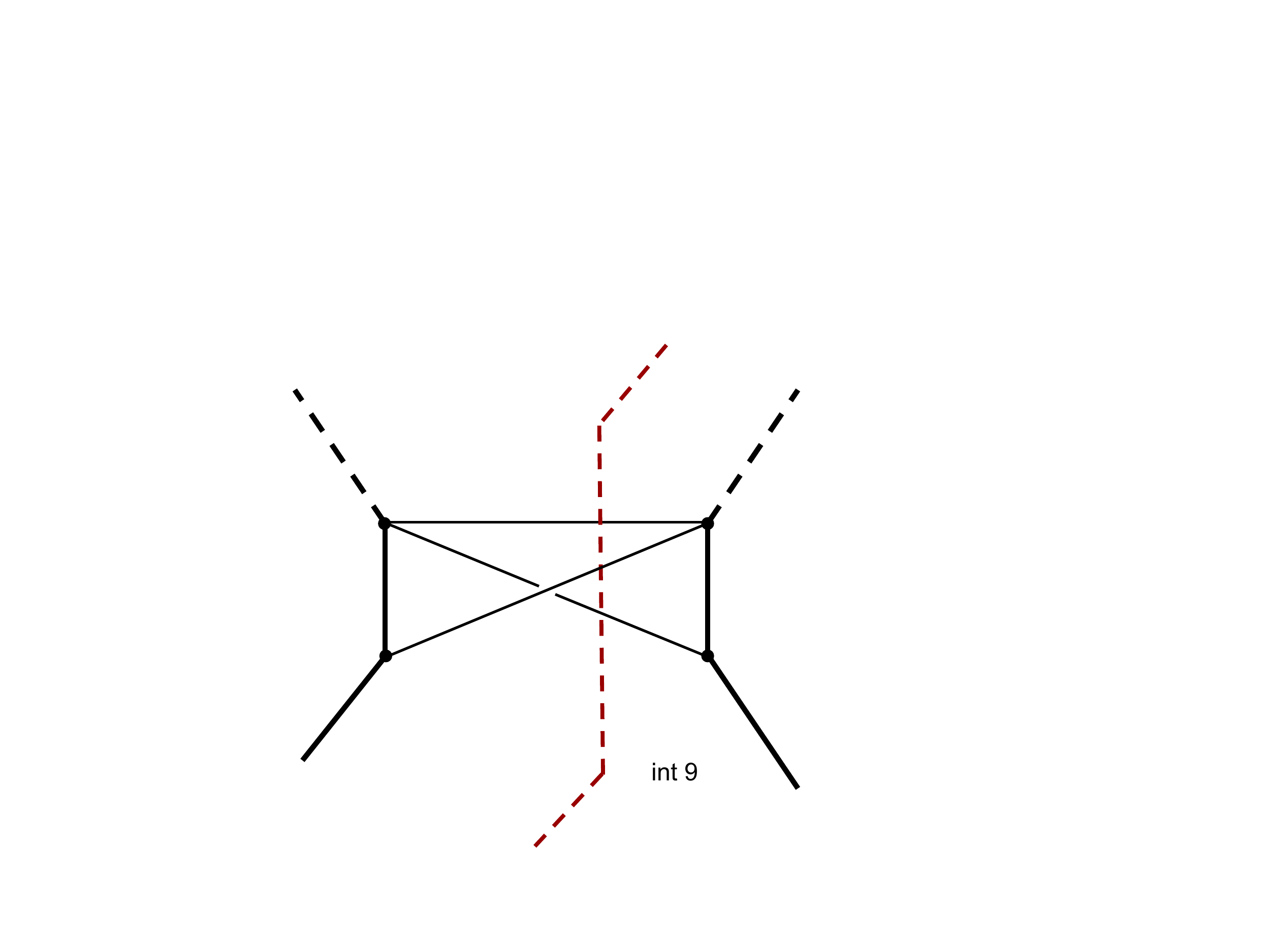} & \hspace*{-4.5cm} \includegraphics[width=11.2cm]{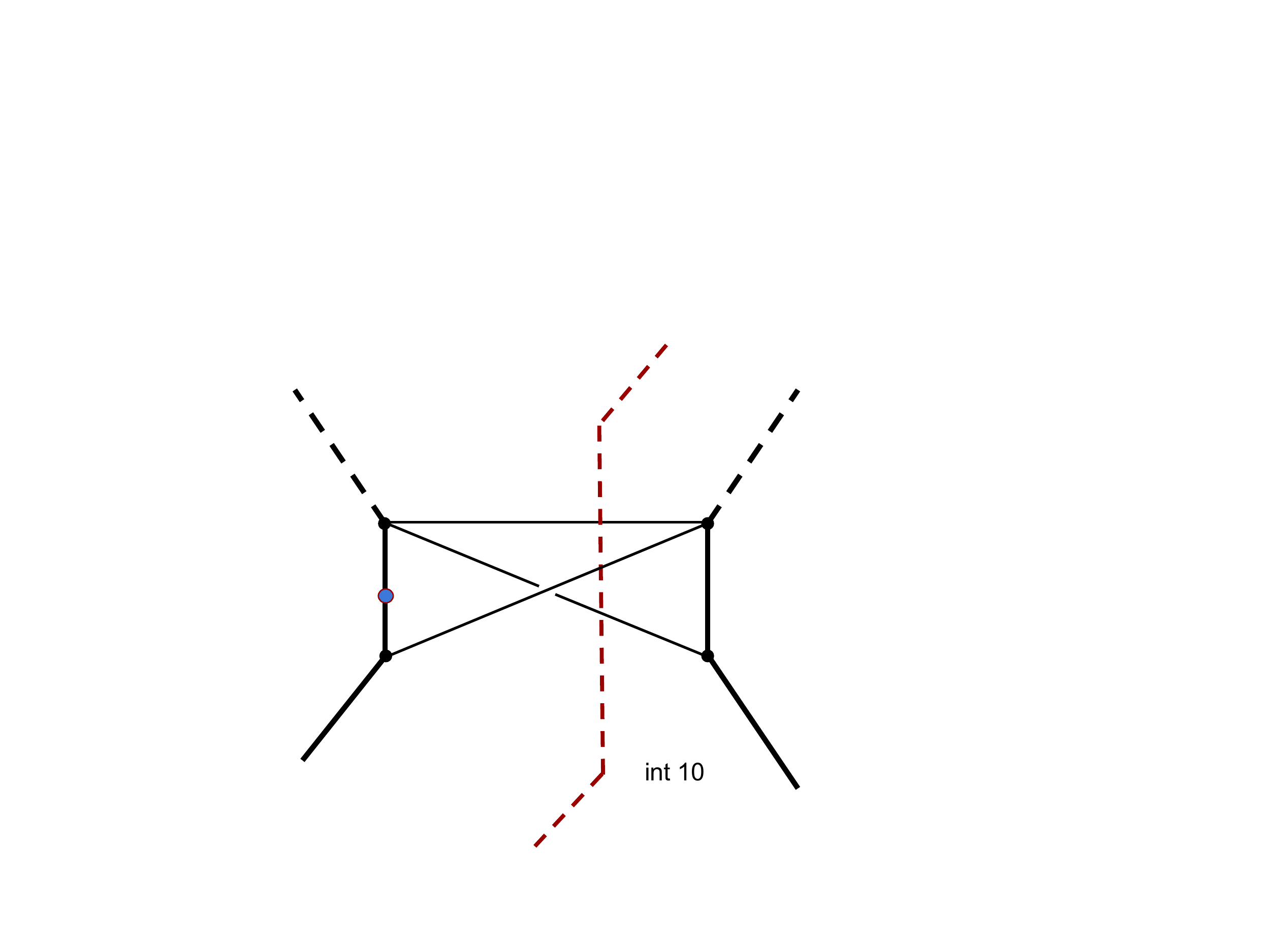} \\
			\vspace*{-4.cm} & \\
			\hspace*{-1.cm}	\includegraphics[width=11.2cm]{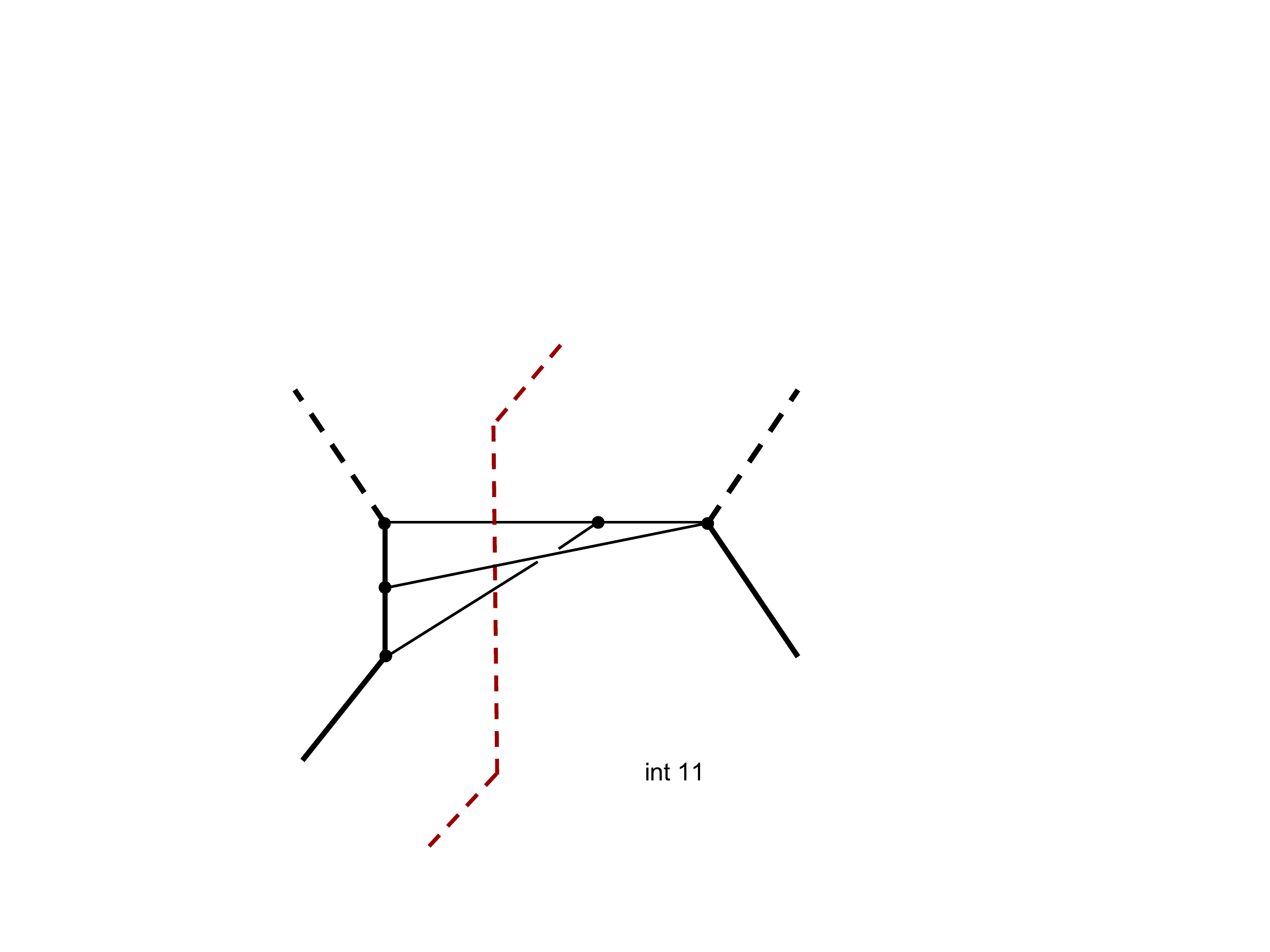} & \hspace*{-4.5cm} \\
		\end{tabular}
		\vspace*{-1cm}
	\end{center}
	\caption{Additional Master Integrals from Topology B; Integrals $I_9,I_{10}, I_{11}$. \label{fig:TopBMIS}}
	
\end{figure}
 
The integrals belonging to topology B are defined in analogy to  Eqs.~(\ref{eq:topA},\ref{eq:propA}):
\begin{equation}
I_B \left(\alpha_1, \alpha_2, \alpha_3, \alpha_4,\alpha_5, \alpha_6, \alpha_7 \right) = \int \frac{d^d k_1}{(2 \pi)^d} \frac{d^d k_2}{(2 \pi)^d} \prod_{i=1}^7  \frac{1}{Q_i^{\alpha_i}} \, ,
\label{eq:topB}
\end{equation}
with
\begin{align}
Q_1 &=  P_2  \, , \quad
Q_2 = (p_1 + k_2)^2 -m_b^2  \, , \quad
Q_3 =P_3 \, ,\quad
Q_4 = (p_2 + k_2)^2 \, , \nonumber \\
Q_5 &= P_5\, ,\qquad
Q_6 = P_6 \, \qquad
Q_7 = P_7 \, ,
\label{eq:propB}
\end{align}
where again $Q_5,Q_6,Q_7$ are cut propagators.
Topology B involves eleven MIs, which include the eight MIs already needed for topology A, plus the three non planar MIs shown in Figure~\ref{fig:TopBMIS}. The latter are labeled as follows
\begin{align}
I_9 \equiv I_B \left(0,1,1,0,1,1,1 \right) \, ,&\qquad   &I_{10} \equiv I_B \left(0,2,1,0,1,1,1 \right) \, , \nonumber \\
I_{11} \equiv I_B \left(1,0,1,1,1,1,1 \right) \, .&\qquad   &
\end{align}

\begin{figure}[htb]
	\vspace*{-2cm}
	\begin{center}
		\hspace*{1cm}
		\includegraphics[width=11.2cm]{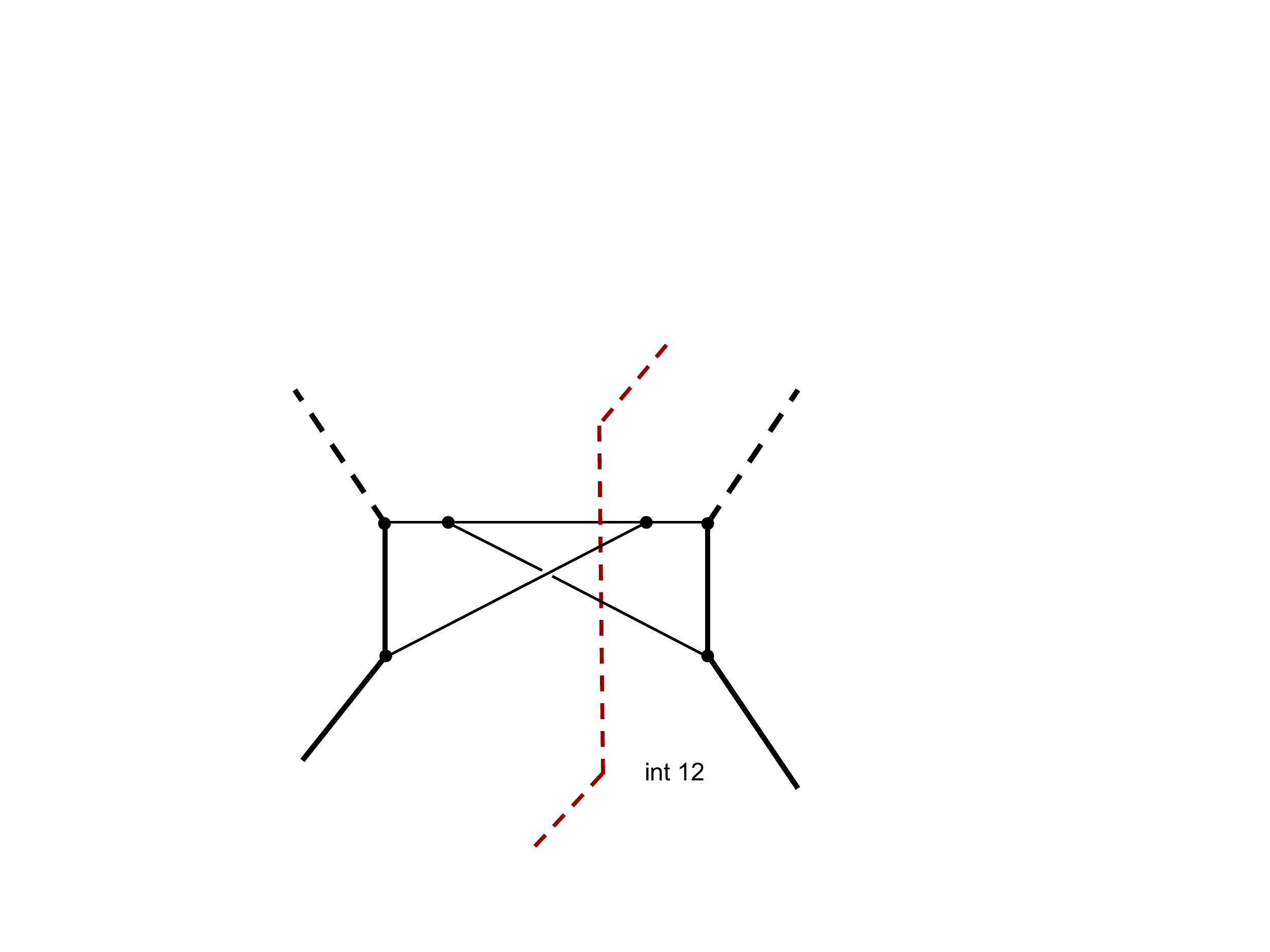} 
	\end{center}
	\caption{Additional Master Integral from Topology C: Integral $I_{12}$. \label{fig:TopCMIS}}
	
\end{figure}

Topology C is defined by the integrals
\begin{equation}
I_C \left(\alpha_1, \alpha_2, \alpha_3, \alpha_4,\alpha_5, \alpha_6, \alpha_7 \right) = \int \frac{d^d k_1}{(2 \pi)^d} \frac{d^d k_2}{(2 \pi)^d} \prod_{i=1}^7  \frac{1}{R_i^{\alpha_i}} \, ,
\label{eq:topC}
\end{equation}
with
\begin{align}
R_1 &=  Q_2  \, , \quad
R_2 = P_3  \, , \quad
R_3 =Q_4 \, ,\quad
R_4 = P_4 \, , \nonumber \\
R_5 &= P_5\, ,\qquad
R_6 = P_6 \, \qquad
R_7 = P_7 \, ,
\label{eq:propC}
\end{align}
Topology C involves five MIs: $I_1$ and $I_4$, which are present also in topologies A and B, $I_9$ and $I_{10}$, which are already needed for topology B and one additional non-planar integral
\begin{equation}
I_{12} \equiv I_C \left(1,1,1,1,1,1,1 \right) \, ,
\end{equation}
which is shown in Figure~\ref{fig:TopCMIS}.

\subsection{Differential equations}

\begin{table}[t]
	\renewcommand*{\arraystretch}{1.5}
	\begin{center}
		\begin{tabular}{|c||c|c|c|c|c|c|c||c|}
			\hline
			Integrals & $1/\epsilon^2$ & $1/\epsilon$ & $\epsilon^0$ & $\epsilon$ &  $\epsilon^2$ &$\epsilon^3$ &$\epsilon^4$ & Numerical Constants\\
			\hline
			\hline
			$I_1$ & -  & - & 0 & 1 &2  & 3 & 4 & none  \\
			\hline
			$I_2$ & - & - & 1 & 2 & 3 & 4  &  X & $C_1,C_2,C_3$\\
			\hline
			$I_3$ & - & 1 & 2 & 3 & 4 & X  & X & $C_1,C_2$\\
			\hline
			$I_4$ & - & - & 1 & 2 & 3 & 4  &  X & none \\
			\hline
			$I_5$ & - & - & 3 & 4 & X & X  &  X & $C_1,C_4,C_5$\\
			\hline
			$I_6$ & - & 1 & 2 & 3 & 4 & X  &  X & $C_1,C_2,C_4,C_5$\\
			\hline
			$I_7$ & - & 0 & 1 & 2 & 3 & 4  &  X& $C_1,C_2,C_4,C_5,C_6$ \\
			\hline
			$I_8$ & - & 2 & 3 & 4 & X & X  &  X& $C_1,C_2$ \\
			\hline
			$I_9$ & - & - & 2 & 3 & 4 & X  &  X& none \\
			\hline
			$I_{10}$ & - & 1 & 2 & 3 & 4 & X  &  X & none \\
			\hline
			$I_{11}$ & 1 & 2 & 3 & 4 & X & X  &  X &  $C_1,C_2, C_4, C_5$\\
			\hline
			$I_{12}$ & 1 & 2 & 3 & 4 & X & X  &  X &  none \\
			\hline
		\end{tabular}
	\end{center}
	\caption{This table summarizes the structure of the $\epsilon$ expansion of the various MIs. The numbers in the table indicate the maximum weight of the GPLs found at a given order of the epsilon expansion. Terms involving GPLs of weight five and  higher were not evaluated (except for the case of $I_5$, see text) and are indicated with an X in the table. \label{tab:MIs}}
\end{table}

Each MI satisfies differential equations with respect to the two dimensionless parameters $z$ and $t$ which we use in order to parameterize the phase space. The differential equations can be derived starting from the IBPs and are directly obtained from {\tt LiteRed}. Only the integrals of topology A involving the propagators $P_2,P_3,P_5,P_6,P_7$ are found to involve more than two MIs. Consequently, since the solutions of a system of three first-order differential equations cannot be found with standard methods, the evaluation of the three MIs with the five propagators listed above could be problematic. However, the integrals $I_5,I_6$ and $I_7$, which we chose as the MIs for this set of propagators, satisfy a system of three differential equations in $t$ and $y$ which decouple order by order in $\epsilon$. For this reason, in the case at hand it was possible to evaluate the twelve MIs in the problem without employing a canonical basis \cite{Henn:2013pwa}.

\begin{table}[t]
	\begin{center}
		\renewcommand*{\arraystretch}{1.5}
		\begin{tabular}{|c|c|c|}
			\hline
			Constant & Integral/Order& value \\
			\hline
			\hline
			$C_1$ & $\hat{I}_2, \epsilon$& \begin{tabular}{cc} & $10.239578222392149411675195562805$ \\  $- i \pi$ &                                    $17.545177444479562475337856971665$    \end{tabular}                              \\
			\hline
			$C_2$ & $\hat{I}_2, \epsilon^2$ & \begin{tabular}{cc} & $54.371980398832744511559678484441$ \\
				 $+ i \pi$ & $36.558523291963772395233838229141$ \end{tabular} \\
			\hline
			$C_3$ & $\hat{I}_2, \epsilon^3$ &\begin{tabular}{cc} $-$ & $47.871542821739167668291943163802 $ \\
				$ - i \pi$ & $3.3493397758103340833715632589416$ \end{tabular}\\
			\hline
			$C_4$ & $\hat{I}_5, \epsilon^0$& \begin{tabular}{cc} &
			$332.70825644476215311648744195295$ \\
				$ + i \pi$ & $177.44567822334599921081142309329$ \end{tabular}\\
			\hline
			$C_5$ & $\hat{I}_5, \epsilon$&\begin{tabular}{cc} & 
			$2633.3473713725843348505159328717$ \\
				$ - i \pi$ & $1825.2057515759382789946953393520$ \end{tabular}\\
			\hline
			$C_6$ & $\hat{I}_5, \epsilon^2$ &\begin{tabular}{cc} $-$ &  $4672.8756370231810881261045894874$ \\
				$ + i \pi$ & $5234.4198993443246068516773900101$\end{tabular}\\
			\hline
		\end{tabular}
	\end{center}
	\caption{Values of the constants determined by evaluating the analytic result for the MIs with {\tt GiNaC}.  The condition imposed in order to fix the constants is that integrals $I_2$ and $I_5$ vanish in the $t \to 1$ limit. The second column indicates the MIs used to fix a given constant and  the power $n$ at which the constant first appears as a cofactor of $\epsilon^n$. In the last column one can find the complex value of the constant, which can in principle be determined at arbitrary precision. \label{tab:constants}}
\end{table}

In order to eliminate square-root weights  in the GPLs which appear in the solution of differential equations satisfied by the MIs, we traded the variable $z$ with the variable $y$ defined through the relation
\begin{equation}
z  \equiv \frac{(1+y)^2}{y} \, .
\end{equation}
For $ 0 \le z \le 2$ the variable $y$ is a pure phase which we choose to parameterize as
\begin{equation}
y \equiv e^{i \alpha} \, ,\qquad \frac{\pi}{2} \le \alpha \le \pi \, .
\end{equation} 

The differential equations with respect to $t$ and $y$ satisfied by the MIs are solved order by order in $\epsilon$ using iterated integration. The solutions depend on several integration constants.  Most of these constants can be fixed by imposing the regularity of the MIs in the $t \to 0$ limit. However, a subset of seven constants is left undetermined once the regularity in $t \to 0$ has been required. MIs are in general not regular in the $t \to 1$ limit; indeed, one expects the differential distribution to be singular in the tree-level kinematic limit. The singular part of the distribution in the tree-level limit was evaluated by using SCET. However, all of the MIs which are finite in $\epsilon$ (i.e. $I_1, I_2, I_4, I_5, I_9$) vanish in the $t \to 1$ limit. In particular, the behavior of $I_2$ and $I_5$ in the tree level limit is sufficient to overconstrain the seven remaining constants.
The analytic expression of the MIs in terms of GPLs of argument $y$ and $t$ are very long but they can be evaluated to arbitrary precision by means of the {\tt GiNaC} routines of \cite{Vollinga:2004sn}. Therefore, six of the seven constants were fixed by requiring that integrals $I_2$ and $I_5$ vanish in the $t \to 1$ limit. They are given in numeric form in Table~\ref{tab:constants} with more that thirty significant digits. In general, we evaluated MIs up to the order in the $\epsilon$ expansion where GPLs of weight four first appear in the result, since one does not expect GPLs of weight five to be present in the NNLO differential distributions we are ultimately interested in. (Table~\ref{tab:MIs} summarizes the order in $\epsilon$  at which the various MIs were evaluated.) A notable exception is represented by the MI $I_5$. Indeed, one of the integration constants which appears alongside GPLs of weight four in $I_7$, appears only at order 
$\epsilon^2$ in $I_5$. The term of $\epsilon^2$ in $I_5$ also involves GPLs of weight five. This is not surprising since the set of MIs that was chosen does not have uniform transcendentality.
Therefore, we evaluated $I_5$ up to order $\epsilon^2$ and required that it vanishes in the $t \to 1$ limit in order to fix this residual constant.
In the following section we present analytic results for the MIs.

\section{Results}
\label{sec:results}

\subsection{Alphabet}

The analytic expressions for the MIs which we evaluated are written in terms of Harmonic Polylogarithms  \cite{Remiddi:1999ew} and (two-dimensional) GPLs \cite{Goncharov:1998kja, Goncharov:2001iea, Gehrmann:2000zt, Vollinga:2004sn} of arguments $t$ and $y$. 
GPLs can be defined recursively, for $n \geq 0$, via the iterated integral
\begin{equation}\label{eq:MultPolyLogdef}
G(a_1,\ldots,a_n;z)=\,\int_0^z\,{dt\over t-a_1}\,G(a_2,\ldots,a_n;t)\,,\\
\end{equation}
for a generic argument $z$ and weights $\{a_1, \cdots, a_n\}$,  assuming $G(z) = G(;z)=1$. The case in which $a_i = 0$ for all $i$ needs to be considered separately:
\begin{equation}\label{eq:all0}
G(0,\ldots,0;z) \equiv \, \frac{1}{n !} \ln^n z\,.\\
\end{equation}
The weights of the GPLs or argument $t$ can depend on $y$. For convenience we define the combinations
\begin{equation}
w_1 \equiv  \frac{1-y}{1+y} \, , \qquad w_2 \equiv \frac{1+y^2}{(1+y)^2} = \frac{z-2}{z} \, .
\end{equation}
Consequently, $w_2$ is real while $w_1$ is imaginary.

The list of weights, i.e. the ``alphabet'', appearing in the GPLs of argument $t$ is
\begin{equation}
\left\{0,\pm 1, \pm w_1, \pm w_2 \right\} \, .
\end{equation} 
The alphabet for the GPLs of argument $y$ includes the weights
\begin{equation}
\left\{0,\pm 1, \pm i \right\}\, .
\end{equation} 

\FloatBarrier

\subsection{Integrals}

The simplest MI in the list is the two-loop phase space diagram shown in the top left panel of Figure~\ref{fig:TopAMIS}, for which one can find an expression which is exact in $d = 4 - 2 \epsilon$ dimensions \cite{Gehrmann-DeRidder:2003pne}
\begin{align}
I_1 &= 2^{-9+8 \epsilon} \pi^{-3 + 2 \epsilon} \frac{\Gamma^3(1-\epsilon)}{\Gamma(2 - 2 \epsilon)\Gamma(3 - 3 \epsilon)} m_b^2 \left(\frac{m_b^2}{\mu^2}\right)^{-2 \epsilon} \left[ z^2 (1-t^2) \right]^{1-2 \epsilon} \, , \nonumber \\
&\equiv m_b^2 K(\epsilon) \left(\frac{m_b^2}{\mu^2}\right)^{-2 \epsilon} \hat{I}_1 \left(z,t,\epsilon\right)  \, , \label{eq:I1}
\end{align}
with 
\begin{equation}
K(\epsilon) = 2^{-9+8 \epsilon} \pi^{-3 + 2 \epsilon} \frac{\Gamma^3(1-\epsilon)}{\Gamma(2 - 2 \epsilon)\Gamma(3 - 3 \epsilon)} \, .
\end{equation}
 The first two terms of the $\epsilon$ expansion of $\hat{I}_1$ read
\begin{equation}
\hat{I}_1 = z^2 (1-t^2) - 2 \epsilon z^2 \left(1-t^2 \right)
\left[G(-1,t)+ G(1,t) - 2 G(0,y) + 4 G(-1,y) \right] + {\mathcal O}(\epsilon^2)\, .
\end{equation}

In analogy with Eq~(\ref{eq:I1}), we introduce the notation
\begin{equation}
I_i \equiv m_b^2 K(\epsilon) \left(\frac{m_b^2}{\mu^2}\right)^{-2 \epsilon} \hat{I}_i \, , \qquad i \in \{2, \cdots, 12\} \, .
\end{equation}
The analytic expression of all $\hat{I}_i$ up to terms involving GPLs of weight four can be found in the ancillary file {\tt MasterIntegrals.txt}.

\begin{figure}[tp]
	\begin{center}
		\begin{tabular}{cc}
			\vspace*{-.5cm} & \\
			\hspace*{-1cm}	\includegraphics[width=7.2cm]{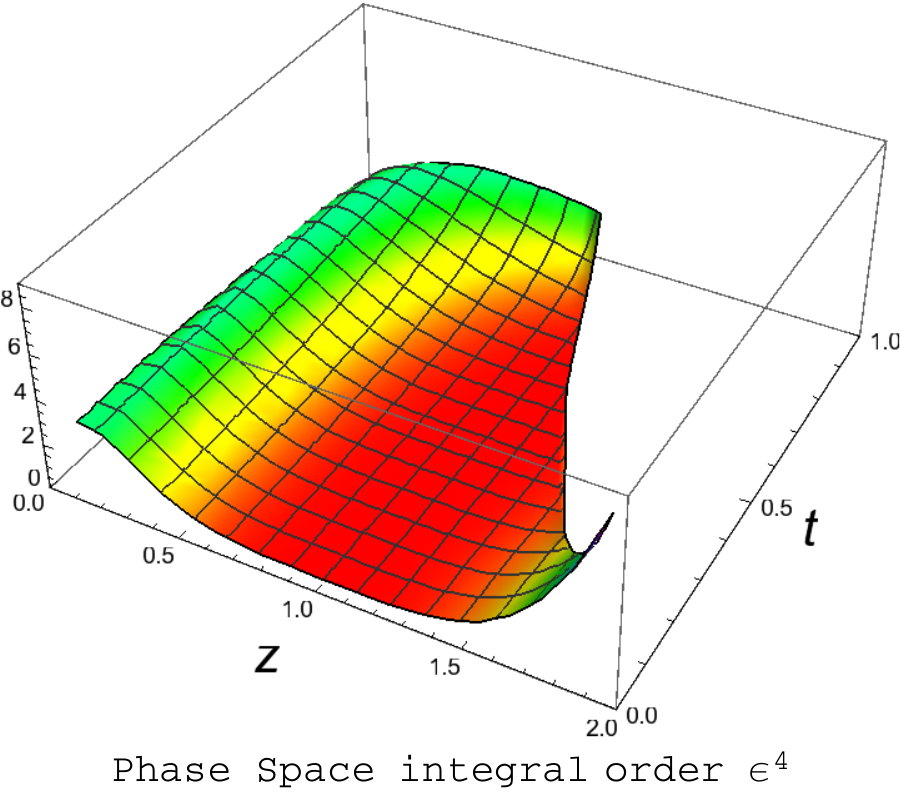} & \hspace*{-0.5cm} \includegraphics[width=7.2cm]{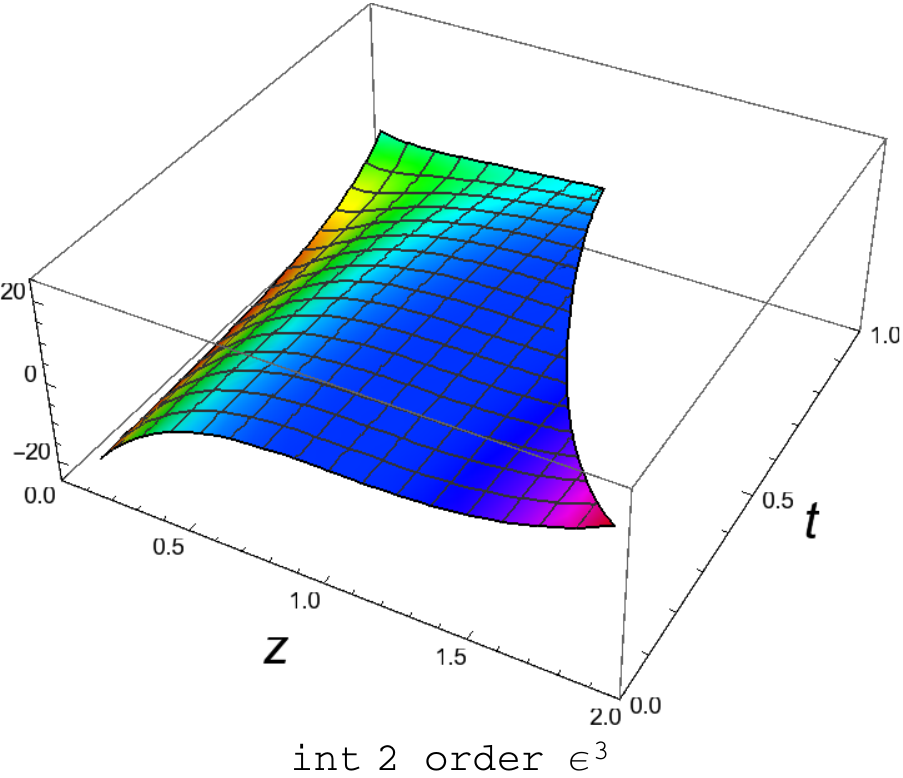} \\
			\vspace*{-0.5cm} & \\
			\hspace*{-1cm}	\includegraphics[width=7.2cm]{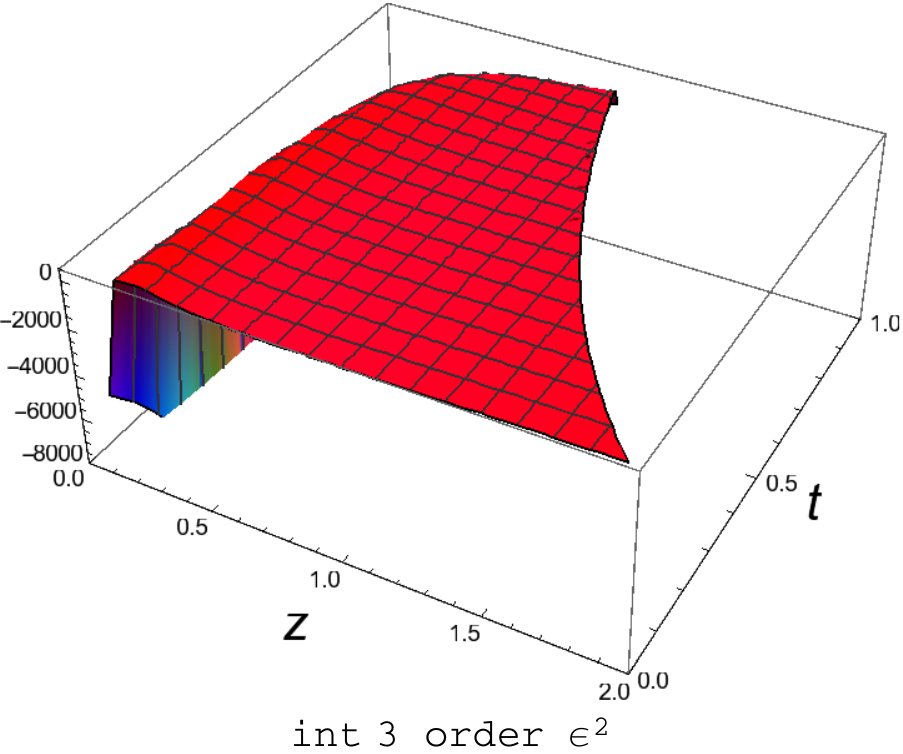} & \hspace*{-0.5cm} \includegraphics[width=7.2cm]{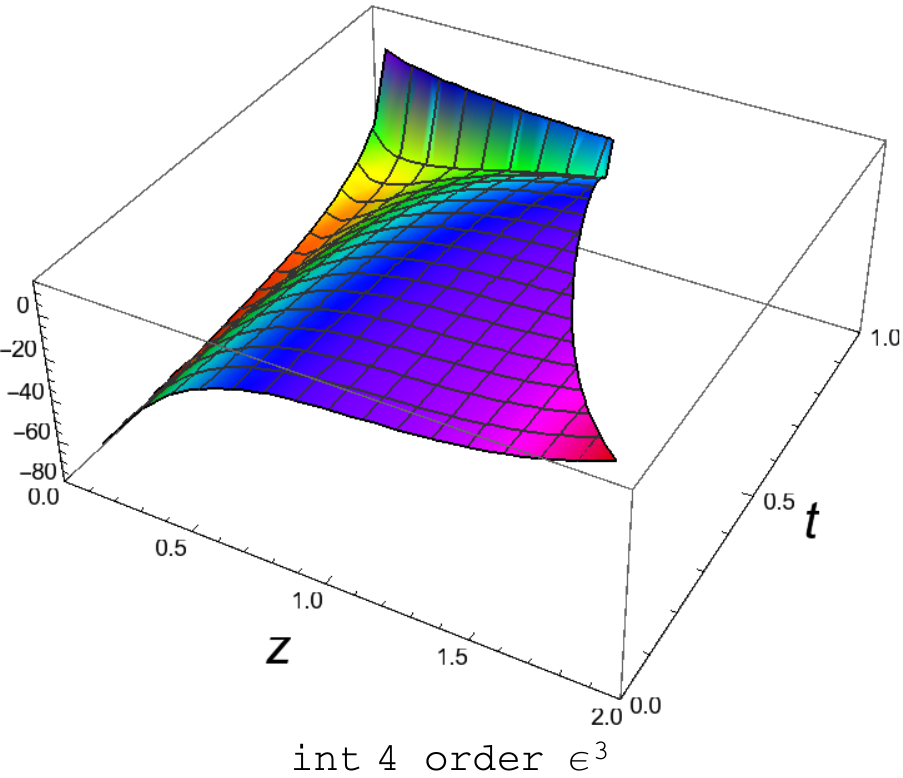} \\
			\vspace*{-0.5cm} & \\
			\hspace*{-1cm}	\includegraphics[width=7.2cm]{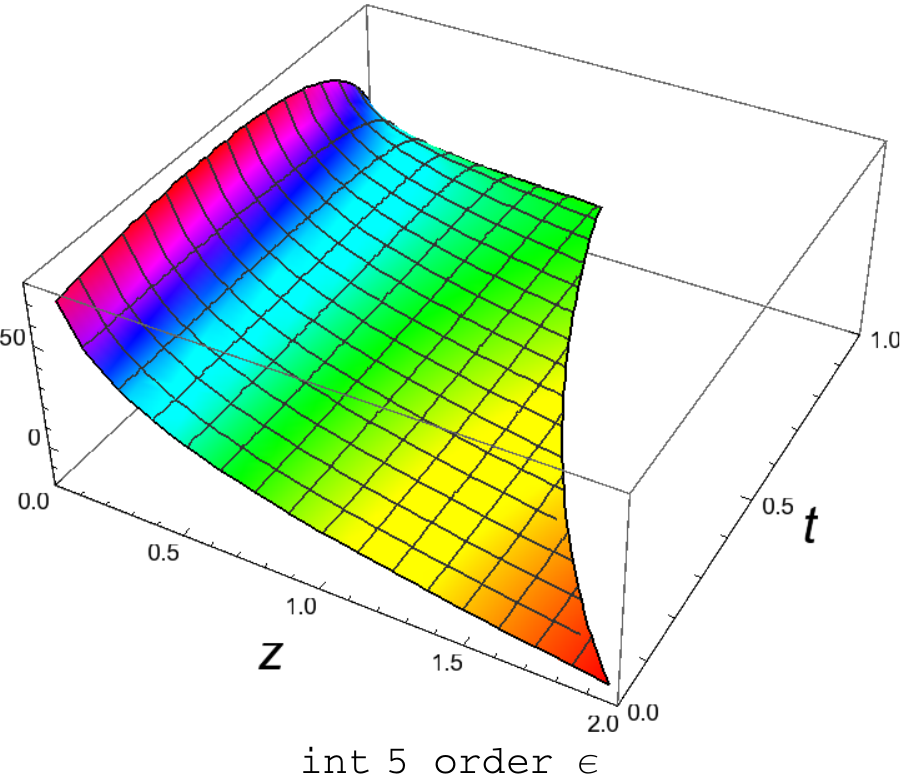} & \hspace*{-0.5cm} \includegraphics[width=7.2cm]{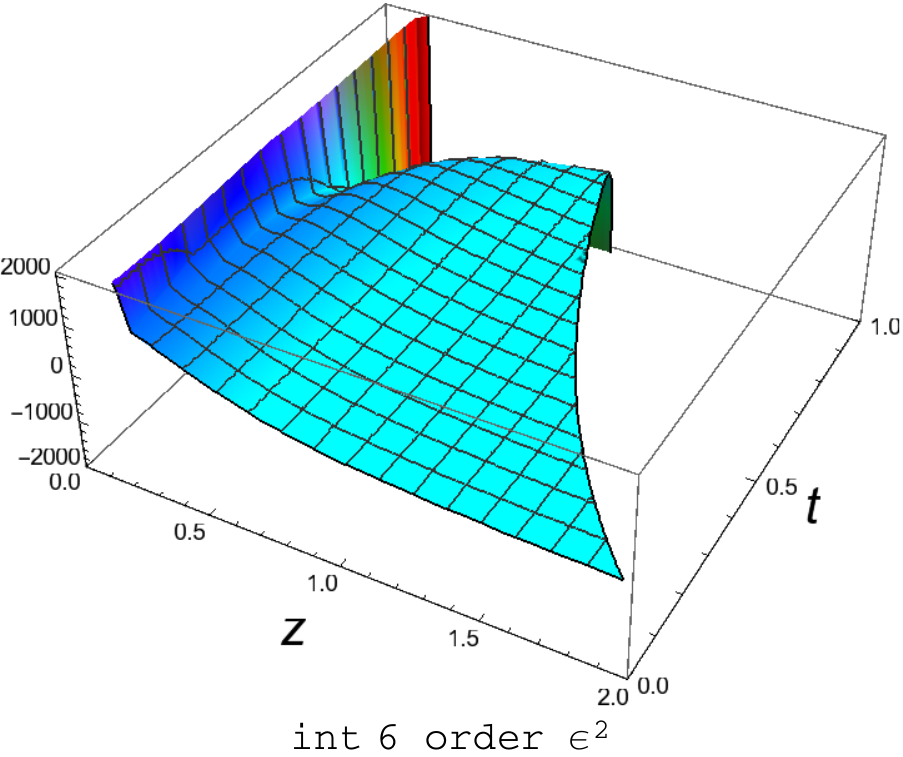} 
		\end{tabular}
	\end{center}
	\caption{Plots of the first order in $\epsilon$ which contains GPLs of weight four for the MIs $\hat{I}_1, \cdots, \hat{I}_6$. \label{fig:TopAhighestorderA}}
	
\end{figure}

\begin{figure}[tp]
	\begin{center}
		\begin{tabular}{cc}
			\vspace*{-.5cm} & \\
			\hspace*{-1cm}	\includegraphics[width=7.2cm]{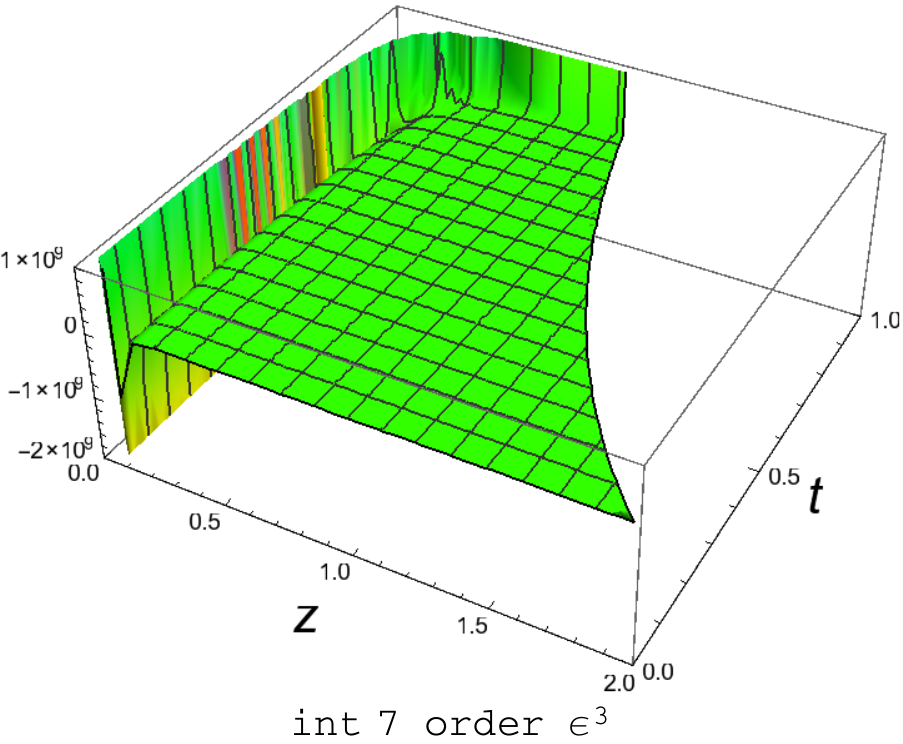} & \hspace*{-0.5cm} \includegraphics[width=7.2cm]{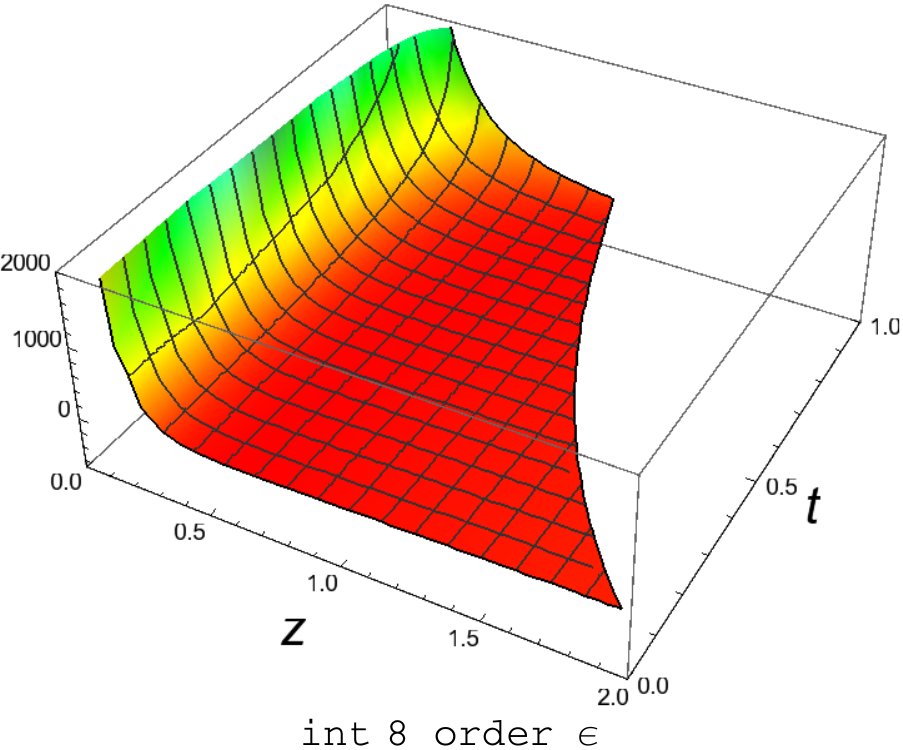} \\
			\vspace*{-0.5cm} & \\
			\hspace*{-1cm}	\includegraphics[width=7.2cm]{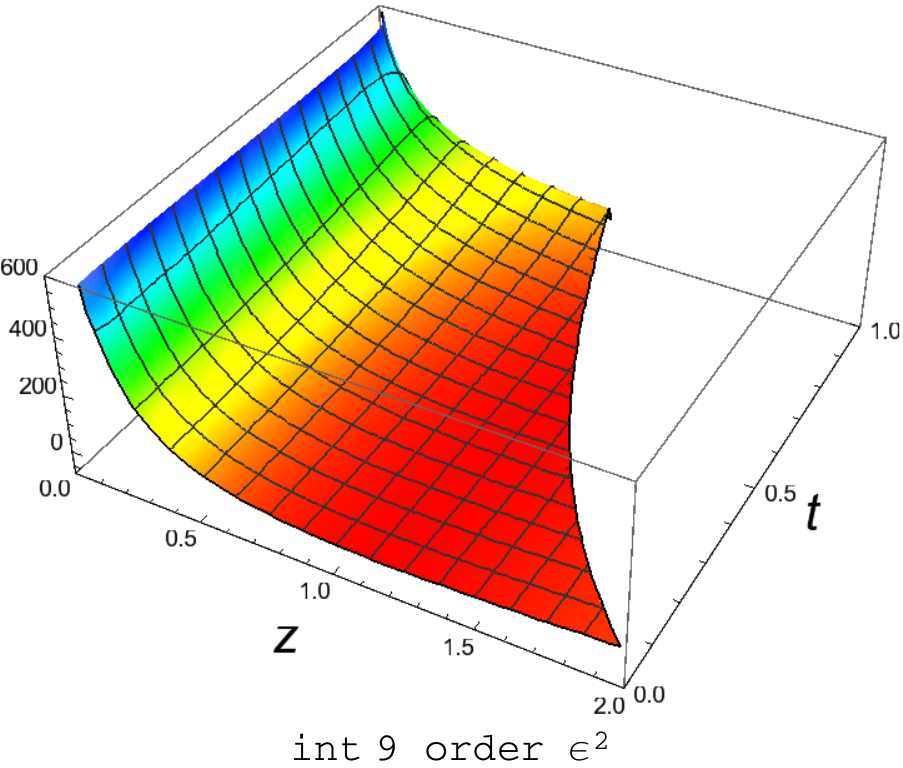} & \hspace*{-0.5cm} \includegraphics[width=7.2cm]{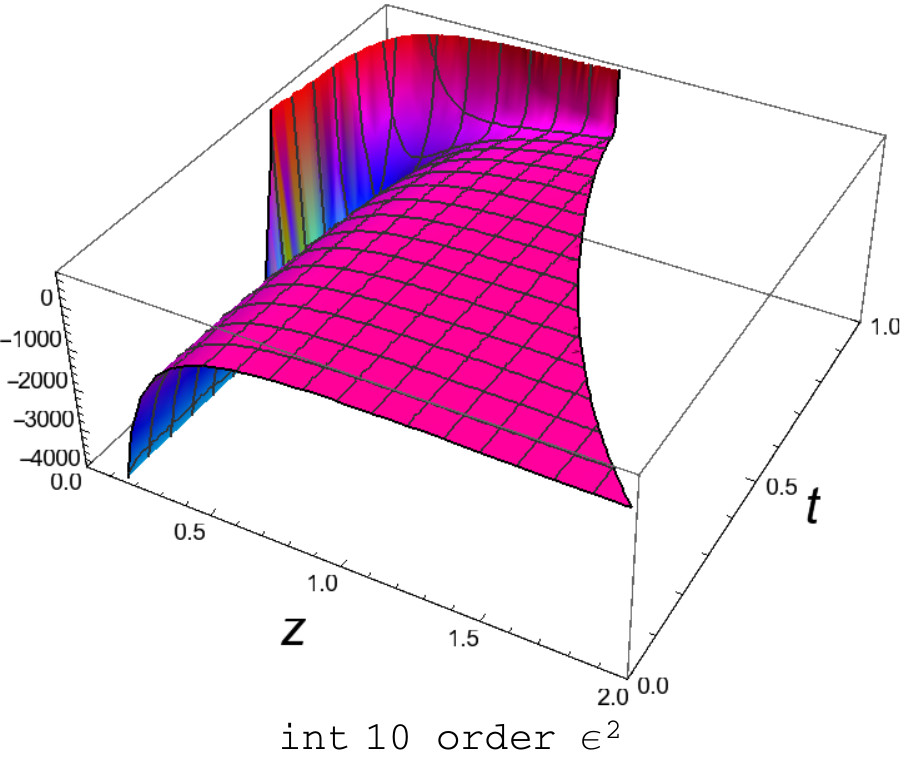} \\
			\vspace*{-0.5cm} & \\
			\hspace*{-1cm}	\includegraphics[width=7.2cm]{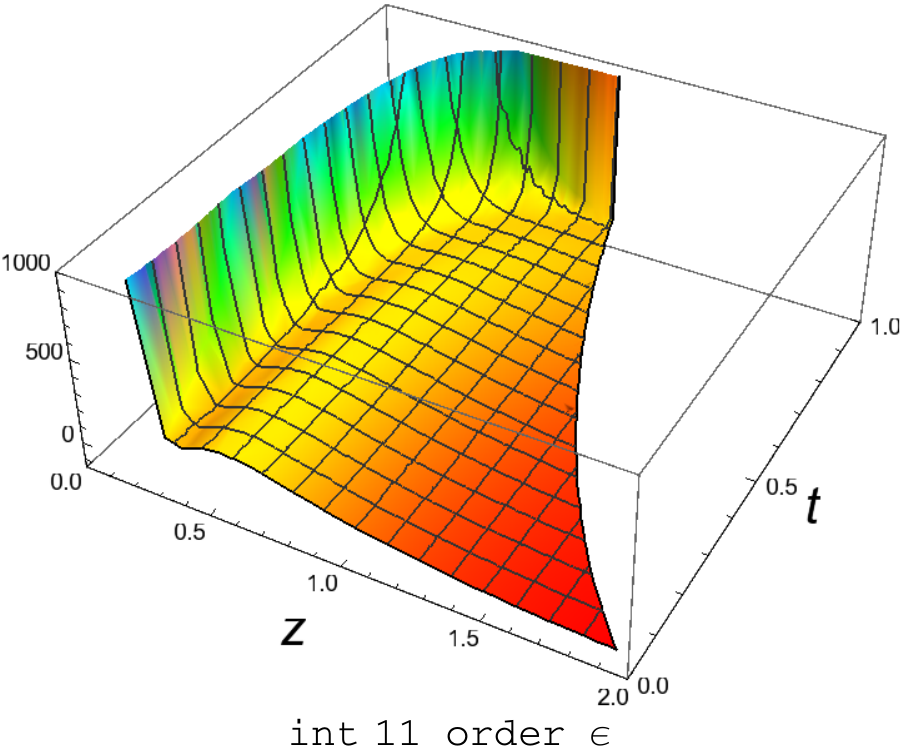} & \hspace*{-0.5cm} \includegraphics[width=7.2cm]{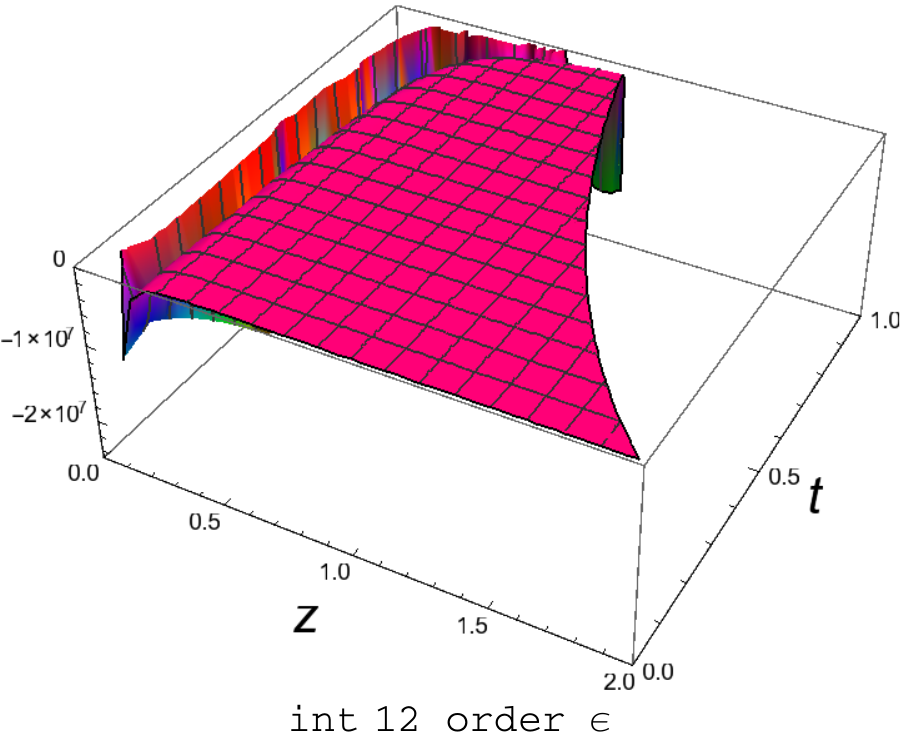} \\
			\vspace*{-0.5cm} & \\
		\end{tabular}
	\end{center}
	\caption{Plots of the first order in $\epsilon$ which contains GPLs of weight four for the MIs $\hat{I}_7, \cdots, \hat{I}_{12}$. \label{fig:TopBhighestorderB}}

\end{figure}

We cross-checked nine out of the twelve MIs which we calculated analytically in this work by comparing the numerical evaluation of their analytic expressions (carried out by means of {\tt GiNaC}) to the direct numerical integration of the MIs, carried out with the package {\tt SecDec} \cite{Carter:2010hi,Borowka:2015mxa,Borowka:2017idc}. We found agreement within the {\tt SecDec} numerical integration error in all points tested. The remaining three MIs ($I_{8}, I_{11}, I_{12}$) belong to subtopologies which also admit at least one two-particle cut. Consequently, these MIs cannot be evaluated directly by comparing them to the imaginary part of a $2 \to 2$ forward box calculated with {\tt SecDec}, since that imaginary part receives a contribution from two-particle and three-particle cuts. Once the MIs corresponding to the two-particle cuts are known, they can be combined with the three-particle cuts calculated here and the sum of two- and three-particle cuts can finally be compared with the imaginary part of the corresponding $2 \to 2$ box integrals.

Finally, we employed {\tt GiNaC} to evaluate the analytic results we found in all of the phase space. In Figures~\ref{fig:TopAhighestorderA} and \ref{fig:TopBhighestorderB} we plot, for all of the integrals $\hat{I}_i$, the  order in $\epsilon$ at which GPLs of order four appear first.

\section{Conclusions}
\label{sec:conclusions}

In this paper we evaluated analytically the MIs needed for the calculation of the double emission corrections to the $b \to u W^*$ decay at tree level. 
The problem was mapped into the calculation of three-particle cuts in two-loop $bW^* \to bW^*$ forward box diagrams.
We identified a set of twelve MIs belonging to three auxiliary topologies by using IBPs. The MIs depend on two dimensionless parameters. Their analytic expression in terms of GPLs was found by means of the DE method. The integrals can be evaluated with arbitrary numerical precision by means of the GPLs functions implemented in {\tt GiNaC}. The complete analytic expression of all of the MIs can be found in an ancillary file included in the {\tt arXiv} version of this paper. The result were cross checked against a direct numerical integration of the MIs carried out by means of the program {\tt SecDec}. The results obtained 
here are needed for the analytic evaluation of the  $b \to u W^*$ decay double differential distribution to NNLO in QCD. Since the two-loop virtual corrections to the $b \to u W^*$ decay are already known analytically \cite{Bonciani:2008wf,Bell:2008ws,Beneke:2008ei,Asatrian:2008uk}, the only missing element at this stage are the one-loop, single-emission diagrams contributing to $b \to u W^*$ decay to NNLO.

\appendix

\section{MIs Poles}
\label{sec:poles}

In this Appendix we collect the explicit expression of the poles of the  MIs which are divergent
in the $\epsilon \to 0$ limit.

\begin{equation}
\hat{I}_3 = - \frac{1}{m_b^4 \epsilon} \frac{8}{z t}  \left[G(-1;t)-G(1;t)\right] + {\mathcal O} \left( \epsilon^0 \right) \, .
\end{equation}

\begin{align}
\hat{I}_6 &= \frac{1}{m_b^6 \epsilon} \frac{32}{t z^2 \left[ 4 + z (t^2-1)\right] }  \left[ G(-1;t) - G(1;t) \right] + {\mathcal O} \left( \epsilon^0 \right) \, , \nonumber \\
\hat{I}_7 &=\frac{1}{m_b^8 \epsilon} \frac{128}{z^4 (t^2-1)^2} + {\mathcal O} \left( \epsilon^0 \right) \, . 
\end{align}

\begin{align}
\hat{I}_8 =& - \frac{1}{m_b^4 \epsilon} \frac{8}{z t} \Bigl[G(-w_2,-w_1;t)+G(-w_2,w_1;t)-G(w_2,-w_1;t)-G(w_2,w_1;t) \nonumber \\
&-G(-1,-w_1;t)-G(-1,w_1;t)+G(1,-w_1;t)+G(1,w_1;t)+G(-w_1,-1;t)\nonumber \\
&-G(-w_1,1;t)+G(w_1,-1;t)-G(w_1,1;t)+2 G(1;y)G(-w_2;t)-2 G(1;y) G(w_2;t)\nonumber \\
&-G(0;y) \left(G(1;t)+G(-w_2;t)-G(w_2;t)\right)-G(-w_2,-1;t)+G(w_2,1;t)\nonumber \\
&+\ln (2)\left( G(w_2;t) - G(-w_2;t) \right) +G(-1;t) \left(G(0;y)-2
G(1;y)+\ln (2) \right) \nonumber \\
&+2 G(1;t) G(1;y)+G(-1,1;t)-G(1,-1;t)-\ln (2) G(1;t)\Bigr] + {\mathcal O} \left( \epsilon^0 \right) \, .
\end{align}

\begin{equation}
\hat{I}_{10} = \frac{1}{m_b^6 \epsilon} \frac{4}{t z} \left[ G(-1;t) - G(1;t) \right]+ {\mathcal O} \left( \epsilon^0 \right) \,.
\end{equation}

\begin{align}
\hat{I}_{11} =& \frac{1}{m_b^6 \epsilon^2} \frac{8}{t z^2 }  \left[ G(-1;t) - G(1;t) \right]+ \frac{1}{m_b^6 \epsilon}\frac{4}{t z^2 } \Bigl\{ 2 \bigl[2 G(-1,-w_1;t)+2 G(-1,w_1;t) \nonumber \\
&-2
G(1,-w_1;t)-2 G(1,w_1;t)+2
G(-w_1,-1;t)-2 G(-w_1,1;t) \nonumber \\
&+2
G(w_1,-1;t)-2 G(w_1,1;t)-5
G(-1,-1;t)-G(-1,1;t)+2 G(0,-1;t) \nonumber \\
&-2
G(0,1;t)+G(1,-1;t)+5 G(1,1;t) \bigr]-\left( G(-1;t) - G(1;t) \right)\bigl[16 G(-1;y) \nonumber \\
&-4 G(0;y)-8 G(1;y)+13+4 \ln (2) \bigr]
\Bigr\} + {\mathcal O} \left( \epsilon^0 \right) \,.
\end{align}

\begin{align}
\hat{I}_{12} =& \frac{1}{m_b^8 \epsilon^2} \frac{128}{t (1-t^2) z^4 } \left[G(-1;t) - G(1;t)\right] + \frac{1}{m_b^8 \epsilon} \frac{64}{t (1-t^2) z^4 } \Bigl\{9 G(1,1;t)-9 G(-1,-1;t)
\nonumber \\
&+G(-1,1;t)+4 G(0,-1;t)-4
G(0,1;t)-G(1,-1;t)+
\left(G(-1;t) -G(1;t) \right) \nonumber \\
&\times \left[8 G(0;y)-16 G(-1;y)-13\right]
\Bigr\}+ {\mathcal O} \left( \epsilon^0 \right) \,.
\end{align}

\section*{Acknowledgments}
The work of A.F. is supported in part by the National Science Foundation under Grant No. PHY-1417354 and PSC CUNY Research Award TRADA-48-265. A.B. and A.F. would like to thank G. Heinrich and S. Jahn for the help with the program {\tt SecDec}. A.B. and A.F. would like to thank the Munich Institute for Astro- and Particle Physics (MIAPP) of the DFG
cluster of excellence ``Origin and Structure of the Universe'' for hospitality and support. 

\FloatBarrier

\bibliography{mybib}

\bibliographystyle{JHEP}

\end{document}